\documentclass[12pt,preprint]{aastex}
\newcommand{\heone}{He~{\footnotesize{I}}}  	% He I
\newcommand{\cartwostar}{C~{\footnotesize{II}}$^{*}$}  	% carbon II*
\newcommand{\carthree}{C~{\footnotesize{III}}}  % carbon III
\newcommand{\nittwo}{N~{\footnotesize{II}}}	% nitrogen II
\newcommand{\oxysix}{O~{\footnotesize{VI}}}  	% oxygen VI
\newcommand{\oxyseven}{O~{\footnotesize{VII}}}  % oxygen VII
\newcommand{\oxyeight}{O~{\footnotesize{VIII}}} % oxygen VIII
\newcommand{\fuse}{{\it{FUSE}}}			% FUSE
\newcommand{\rosat}{{\it{ROSAT}}}		% ROSAT
\shorttitle{O VI halo}
\shortauthors{Shelton et al.}
\author{Robin L. Shelton}
\affil{Department of Physics and Astronomy, the University of Georgia,
        Athens, GA 30602}
\email{rls@hal.physast.uga.edu}
\author{Shauna M. Sallmen}
\affil{Department of Physics, Univ. of Wisconsin - La Crosse, 
La Crosse, WI 54601}
\email{sallmen.shau@uwlax.edu}
\author{Edward B. Jenkins}
\affil{Princeton University Observatory, Princeton, NJ 08544-1001}
\email{ebj@astro.princeton.edu}

\begin{document}
\title{The Galactic Halo's \oxysix\ Resonance Line Intensity}

\begin{abstract}

We used \fuse\ to observe ultraviolet 
emission from diffuse \oxysix\ in the
hot gas in the Galactic halo.
By comparing our result with another, nearby
observation blocked by an opaque cloud at a distance of 230~pc, we could
subtract off the contribution from the Local Bubble, leading to an 
apparent halo intensity of
$I_{OVI}$ = $4680^{+570}_{-660}$ photons cm$^{-2}$ s$^{-1}$ sr$^{-1}$.  
A correction for foreground extinction leads to an intrinsic intensity 
that could be as much as twice this value.  Assuming 
$T\sim 3\times 10^5\,$K, 
we conclude that the electron density,
$n_e$, is $0.01-0.02$ cm$^{-3}$, the thermal pressure, $p/k$, is
$7000-10,000\, {\rm cm}^{-3}\,$K, and that
the hot gas is spread over a length of 50-70~pc,
implying 
a small filling factor for \oxysix-rich gas.
\rosat\
observations of emission at 1/4$\,$keV in the same direction indicate that
the X-rays are weaker by a factor of 1.1 to 4.7, depending on the 
foreground
extinction.
Simulated supernova remnants evolving in low density gas have similar
\oxysix\ to X-ray ratios when the remnant plasma is
approaching collisional ioinizational equilibrium
and the physical structures are approaching dynamical ``middle age''. 
Alternatively, the plasma can be described by a temperature power-law.  
Assuming that the material is approximately isobaric and the
length scales according to
$T^\beta d \ln T$, we find $\beta = 1.5\pm0.6$ and an
upper temperature cutoff of $10^{6.6(+0.3,-0.2)}$~K.  The radiative
cooling rate for the hot gas, including that which is too
hot to hold \oxysix, is $6\times 10^{38}\,
{\rm erg~s}^{-1}{\rm kpc}^{-2}$. 
This rate implies that $\sim$70$\%$ of the energy produced in
the disk and halo by SN and pre-SN winds is radiated by the hot gas in
the halo.

\end{abstract}



\section{Introduction}

Ultraviolet and soft X-ray observations by {\it{Copernicus}}, 
{\it{Voyager}}, 
{\it{HUT}}, 
{\it{ORFEUS}}, {\it{FUSE}}, {\it{SPEAR}},
the Wisconsin All Sky Survey, {\it{HEAO}}, {\it{ROSAT}}, {\it{Chandra}}, 
and {\it{XMM}} detected \oxysix\ ions and soft X-ray emission from the
%high ions in the 
interstellar medium above the Milky Way's disk
\citep{jenkins_1978b,murthy_etal,davidsen,hurwitz_bowyer,wakker_etal,
sheltonetal01, dixonetal, korpela_etal,
mccammon_sanders,
snowden_etal_91,yao_wang,breitschwerdt_cox,henley_etal}.
If the gas is in collisional ionizational equilibrium (CIE), 
then the diffuse \oxysix\ ions and their 
1032, 1038 \AA\ resonance line emission trace $3 \times 10^5$~K gas, 
while the 
diffuse 
1/4 keV and 3/4 keV soft X-rays trace 
%1 to 10$\times 10^6$~K gas
$10^6$ to $10^7$~K gas
\citep{mazzotta_etal,raymond_smith}.
It is natural to 
anticipate that the \oxysix-rich and 1/4 keV emitting plasmas are 
causally, and 
perhaps 
physically, associated because million degree gas eventually cools 
into the 
300,000 K regime.  As it cools, its \oxyseven\ ions 
%(observable with {\it{Chandra}} and {\it{XMM}}) 
recombine with electrons to make \oxysix\ ions.
Furthermore, as time goes by, interfaces between $10^6$~K gas and  
warm ionized gas ($\sim10^4$~K) should develop at temperatures
that are between these two values and  exhibit intermediate levels
of ionization. 
Thus, it is reasonable to expect the 1/4~keV emitting gas to be clad 
in an 
\oxysix-rich sheath, irrespective of whether the hot extraplanar gas 
is due to 
fountains, superbubbles, or extraplanar supernova (SN) explosions.
Such bubble-sheath geometries are present in the disk simulations of
\citet{avillez_breitschwerdt}, as are $T \sim 3 \times 10^5$~K bubbles
without $T \sim 10^6$~K cores.

In this paper, we focus on hot gas above the Galactic disk.   
Even though the 
nearest part of this region may be only a few hundred parsecs 
above the 
midplane, we will follow the X-ray community's tradition by 
referring to this 
region as the halo. 
There are a number of fundamental issues that we plan to 
address about the 
character of the material that creates \oxysix\ and X-ray emissions:
1.) Are the source structures in the halo young, middle aged, or ancient?
2.) What is the temperature distribution function of this gas?
and 3.) How long is this plasma's cooling timescale and how 
does the energy loss 
rate compare with the rate of energy injection from supernova 
explosions in both the disk and halo?

Our analyses are based on sets of 
{\it{Far Ultraviolet Spectroscopic Explorer}} (\fuse)
\oxysix\ and {\it{Roentgen Satellite}} (\rosat) 1/4 keV emission data 
for $l=278.7^{\rm{o}},b=-47.1^{\rm{o}}$
and $l=278.6^{\rm{o}},b=-45.3^{\rm{o}}$.
We focus much of our
attention on the O VI emission data, which can be compared to the soft
X-ray emission to synthesize a long baseline spectrum.  In turn, this
spectrum defines the shape of the plasma's temperature distribution
function and can be compared with models of supernova remnant (SNR) 
evolution to
constrain the age of a remnant, on the assumption that the hot gas in
our line of sight is dominated by the products of one such remnant.
%We took \oxysix\ emission data rather than \oxysix\ absorption 
%column density data. 
%Together, the \oxysix\ and soft X-ray measurements comprise a 
%long baseline
%spectrum, which can be used to constrain models.   We use the ratio
%of \oxysix\ to soft X-ray intensity to constrain a model for 
%the time since
%shock heating and to determine the shape of the plasma's temperature
%distribution function.

We have no direct measurement of the \oxysix\ column density along our
line of sight.  Nevertheless, observations of \oxysix\ absorption
features toward bright extragalactic sources at only moderate angular
separations from our direction give an approximate guide on the value of
the \oxysix\ column density, $N_{\rm OVI}$.  This information is
useful, since both \oxysix\ and soft X-ray emission intensities 
are equal to
$(1/4\pi)\int n_en_i<\sigma v>(T)dl$, where $n_e$ is the local 
electron density, 
$n_i$ is the local density of ions, and $<\sigma v>(T)$ is a 
temperature-dependent emissivity function.   
In contrast,
the \oxysix\ column density is simply equal to the line of sight
integral of the ion density.   The column density integral
is not additionally weighted 
by electron density, and its temperature dependence is simply 
that due to the 
changes in the ion fraction.   
As a result, the ratio of the \oxysix\ intensity to column density 
%(taken from the literature)
can be used to determine the electron density and from that, the thermal
pressure.

Every observation of the halo's intensity is contaminated 
by photons produced elsewhere along the line of sight.  The most obvious
non-halo sources are the heliosphere \citep{lallement} and the 
Local Bubble 
(also called the Local Hot Bubble), where the latter is an 
%($\sim10^6$~pc), 
irregularly shaped pocket of X-ray emissive, interstellar gas surrounding 
the Solar neighborhood out to a distance of $\sim$100~pc 
\citep{cox_reynolds}.  
The combined Local Bubble and 
heliospheric emission must be subtracted from that found for observations 
of the high latitude sky in order to determine the halo's intensity.  To 
accomplish this, we selected two adjacent lines of sight, 
one that is nearly 
free of foreground obscuration and another that has a cloud 
of neutral gas, 
which 
%blocks the X-rays from the halo 
blocks the more distant halo emission
but not the foreground emission.  
By evaluating the difference between the two intensities, we can measure 
the emission that arises only from the halo. This procedure is called the 
``shadowing technique.''  

\begin{figure}
%\epsscale{0.7}
%\plotone{../figures/ds9.eps}
\plotone{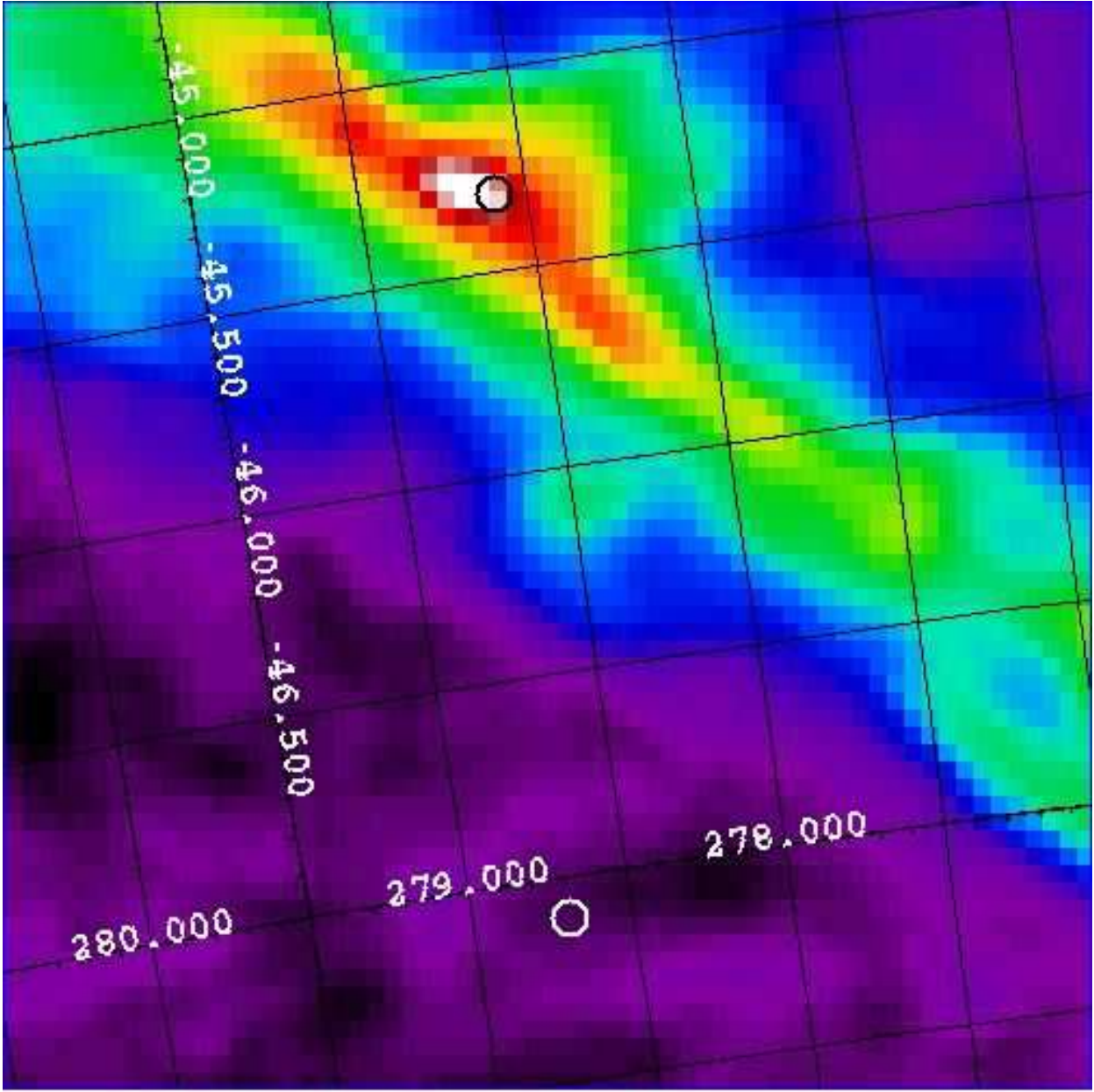}
  \caption{The shadowing filament and nearby sky are shown in
this infrared map made from {\it{DIRBE}} corrected {\it{IRAS}} 
100 $\mu$m data.
The filament runs diagonally across the upper right
portion of the image.
During the ``shadowed'' observation, \fuse\ was
pointed toward the most opaque portion of the filament
($278.6^{\rm{o}},-45.3^{\rm{o}}$, within the upper circle).
During the ``unshadowed'' observation, \fuse\ was pointed
toward a low opacity region just off of the filament 
($278.7^{\rm{o}},-47.1^{\rm{o}}$, within the lower circle).
The \fuse\ \, LWRS aperture measures $30'' \times 30''$, 
thus a small fraction of the size of the overlayed circles.}
\label{fig:diras}
\end{figure}

For our shadowed line of sight, we used the strength of
\oxysix\ emission found along the path to a nearby, opaque filament.
The filament is located $230\pm30$ pc from the Sun \citep{penprase_etal}, 
so is positioned beyond, but not far beyond the Local Bubble.   
\fuse\ \ has already observed a dense part of the filament 
at $l=278.6^{\rm{o}},b=-45.3^{\rm{o}}$.
The observation yielded a 2$\sigma$ upper limit on the 
\oxysix\ doublet's intensity, including the systematic uncertainties,
of 800 photons s$^{-1}$ cm$^{-2}$ sr$^{-1}$ 
\citep{shelton_2003}.    
Ignoring the systematic uncertainties and using the $1\sigma$ statistical
uncertainties as error bars leads to a best fit value of 
% unrounded: $26 \pm {339}$
% rounded:
$30 \pm {340}$ photons s$^{-1}$ cm$^{-2}$ sr$^{-1}$
(as determined by one of us (RLS) while preparing the original article).
%The 1$\sigma$ upper limit, excluding systematic uncertainties,
%% unrounded:  $26^{+339}_{-26}$ photons s$^{-1}$ cm$^{-2}$ sr$^{-1}$ 
%is $30^{+340}_{-30}$ photons s$^{-1}$ cm$^{-2}$ sr$^{-1}$
%(as determined by one of us (RLS) while 
%preparing the original article)). 
%%We take the strength of the local \oxysix\
%%contributions from prior \fuse\ observations of an 
%opaque filament residing 
%%beyond, but not far beyond the Local Bubble 
%(filament's distance from Sun:  
%%$230\pm30$ pc \citep{penprase_etal}; direction: 
%%$l=278.6^{\rm{o}},b=-45.3^{\rm{o}}$; \oxysix\ doublet's 
%%2$\sigma$ upper limit 
%%including $14\%$ systematic uncertainty:
%%800 photons s$^{-1}$ cm$^{-2}$ sr$^{-1}$ \citep{shelton_2003}, 
%%1$\sigma$ upper limit excluding systematic uncertainty: 
%%% unrounded:  $26^{+339}_{-26}$ photons s$^{-1}$ cm$^{-2}$ sr$^{-1}$ 
%%$30^{+340}_{-30}$ photons s$^{-1}$ cm$^{-2}$ sr$^{-1}$ 
%%(as determined by one of us (RLS) 
%while preparing the original article)). 
To measure 
the combined halo and Local Bubble \oxysix\
intensity for this paper, we used \fuse\ to observe an unobscured 
line of sight only 2$^{\rm{o}}$ from the filament 
($l=278.7^{\rm{o}},b=-47.1^{\rm{o}}$; see Figure~\ref{fig:diras}).
In order to estimate
the halo's soft X-ray count rate, we perform a similar shadowing
analysis using \rosat\ data for the same pointing directions.
We also subtract the extragalactic soft X-ray contribution and
account for the variation in optical depth with photon energy.
By carefully isolating the halo emission, we expect to circumvent
the amalgamation problem seen in plots of total \oxysix\ intensity
versus total soft X-ray countrate.

In Section~\ref{sect:observations}
of this paper
we present our measurements of the \oxysix\ emission produced along the
unobscured, off-filament line of sight.
%In Subsection~\ref{subsect:oviintensity},   
%we determine the halo's \oxysix\ intensity
%and in Subsection~\ref{subsect:physicalproperties}, 
%we use it to estimate the 
%density and pressure of the emitting material.
%In Subsection~\ref{subsect:xrayspectrum}, we perform 
%a shadowing analysis using 
%\rosat\ data
%from the filament and off-filament directions and resulting in
%an estimate of the halo's 1/4~keV intensity.   
In Subsection~\ref{subsect:oviintensity}, we subtract the local 
\oxysix\ intensity from our measurement of the unobscured line of
sight, yielding the intensity originating
in the halo.    We determine the halo's \oxysix\ column density
from published data for nearby directions
in Subsection~\ref{NOVI_estimate}.
In Subsection~\ref{subsect:xraybrightness}, 
we estimate of the halo's 1/4~keV surface brightness by performing a 
shadowing analysis with \rosat\ data from the filament and off-filament 
directions.
We then use the halo's \oxysix\ intensity and 
column density determinations 
and treat the \oxysix-rich gas as if it is isothermal in order
to estimate the density and pressure of the emitting material 
in Subsection~\ref{subsect:physicalproperties}.
We compare the \oxysix\ and 
1/4~keV results with simulation predictions 
in order to estimate the plasma's maturity
in Subsections~\ref{subsect:halosnr}.
We compare the \oxysix\ and 1/4~keV results with analytic functions 
in order to estimate the plasma's temperature distribution function,
pressure, and cooling rate in Subsection~\ref{temp_dist}.
The results are summarized in Section~\ref{sect:summary}.

\section{Observations, Data Reduction, Spectral Analysis, 
and Determination of the Halo's O VI Intensity} 
\label{sect:observations}

\subsection{Observations}

\fuse\ observed the blank sky toward
$l=278.69^{\rm{o}},b=-47.08^{\rm{o}}$ 
(R.A. = $03^{\rm{h}} 20^{\rm{m}} 10^{\rm{s}}$, 
decl. = $-62^{\rm{o}} 26' 29''$) once in July and 
twice in November of 2002,
for Guest Observer program C153.  
These observations
exposed the LiF 1A detector segment, the segment used for
the subsequent \oxysix\ signal extraction, for
53, 104, and 61~ksec, respectively.
Of these durations, 
6, 61, and 26~ksec, respectively, occurred while the satellite was 
in the night portion of its orbit.
\fuse\ also made a non-proprietary ``Early Release Observation''
of this direction in December 1999 and archived the
data under program identification number X0270301.
However, detector 1 was de-powered during most of the observation time.

\subsection{Data Reduction}

During most of the Guest Observer exposures, data were taken on all four 
detector segments
(1A, 1B, 2A, and 2B).
The data from each detector segment
were taken in time-tag mode where the photoevents from each detector are 
listed in the sequence that they were recorded, 
with their times and locations on
the detectors where they were found.
%consist of photon lists, which are lists of times when
%individual photons were detected and the locations on the detector
%where they were found.
%(1999, December; 2002, July; and 2002, November),
We concatenated the photon lists obtained from the
individual exposures, creating lists for each detector 
segment and observation.
% The following note about concatenation applied to my 
% old method with the CalFUSE v2.2 data, but not the CalFUSE v3.1+ data.
%The 2002, November viewing was treated as a single observation
%because it was taken in an uninterrupted series of exposures,
%even though it was archived under multiple identification numbers 
%(C1530102 and C1530103).

We then ran the concatenated photon lists through
version 3.1.1 of the CalFUSE pipeline 
\citep{sahnow_etal,datahandbook}, with corrections enabling us
to use the extended source extraction window.
We did not use automatic background subtraction.
For the LiF 1A
extractions, we set the lower and upper
pulse height limits to 4 and 25, respectively.  For the other
detector segments, we used the default pulse height limits.
This process yielded several spectra, because
in addition to recording emission with four detector segments,
\fuse\ records light reflected off of each of
two types of coatings (lithium fluoride (henceforth
called LiF) and silicon carbide (henceforth called SiC))
and four gratings.   The multiplicity 
enables the telescope to see eight overlapping
wavelength regimes with various efficiencies.
Furthermore, each mirror's focal plane assembly has
four apertures.  Thus, each detector segment is able to receive
light from each of the four apertures simultaneously 
and without overlap.  
% the following statement applies only to previous versions of pipeline:  
%By processing
%all of the data through the pipeline, we obtain spectra from 
%all eight channels and all four apertures.  
The spectra observed through the largest
aperture (LWRS: 30'' $\times$ 30'') in the LiF1A channel 
provides the greatest signal to noise and the greatest 
effective area in the 1030 to 1040 \AA\ regime.
We will use these spectra 
for the following \oxysix\ data reduction, and will use the 
spectra from the other channels when searching for other cosmic emission
lines outside of the 1030 to 1040 \AA\ bandpass.  
Our avoidance of the SiC 
channels overcomes the solar \oxysix\ line contamination 
problem discussed by 
\citet{lecavelier_etal}.
%Lecavelier des Etangs, Gopal-Krishna \& Durret (2004).
% above reference is A&A, 421, 503.

The wavelength scales for each spectrum require re-calibration,
which we 
%performed 
determined 
by comparing the reported heliocentric wavelengths of the 
observed geocoronal emission lines 
around 1027 and 1040 \AA\
with their rest wavelengths \citep{morton} converted
from the geocentric to the heliocentric reference frame.
We fit a linear function of wavelength to the difference between the 
expected and observed heliocentric wavelengths, then subtracted this
function from the observed wavelengths.
We then combined like spectra 
and shifted to the Local Standard of Rest (LSR) reference frame.

\subsection{Spectral Analysis}

\begin{figure}
\epsscale{0.7}
%\plotone{../figures/spectra.ps}
\plotone{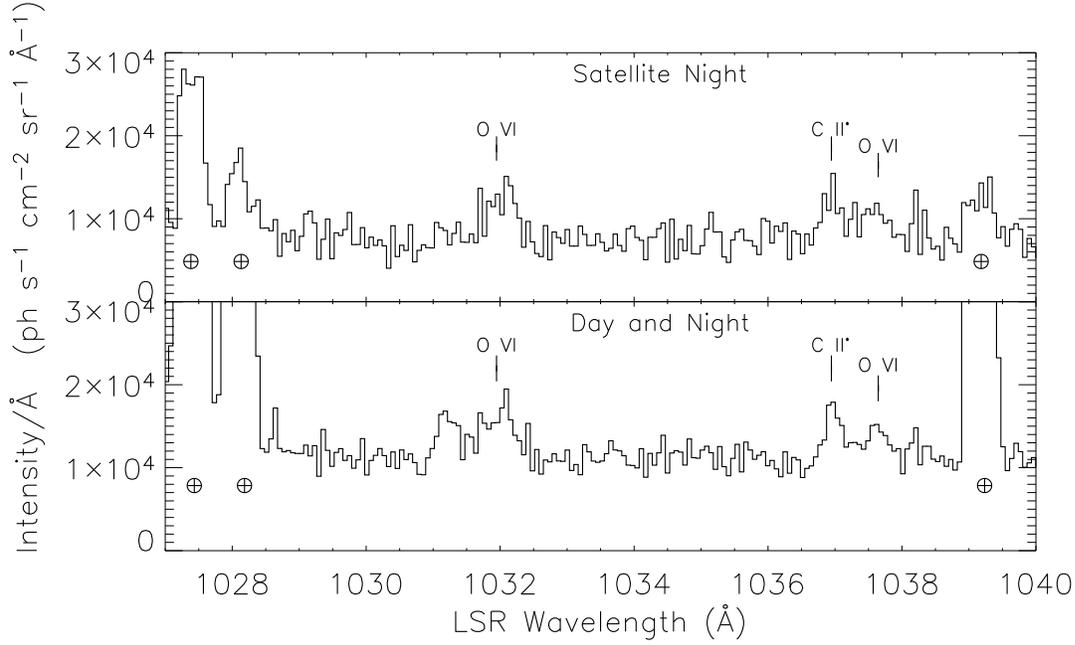}
\vspace{1cm}
  \caption{LiF1A spectra, binned by  0.065 \AA\ 
% (10 old pixels, but Shauna binned by 2 and then I summed by 5)
and plotted relative to
the corrected wavelength for the Local Standard of Rest (LSR) 
reference frame.
Top panel: spectrum from the satellite-night portion of the data.
Bottom panel:  spectrum from the full dataset. 
The \oxysix\ 1032 and 1038 \AA\ emission lines and
the \cartwostar\ emission line appear in both
the night only and the day+night spectra.   
The feature at 1031.2 \AA\ 
in the day+night spectrum 
%is observed in all long \fuse\ blank sky observations and 
may be the second order diffraction of an atmospheric emission
feature of 
\heone\ at 515.62$\,$\AA.
% (first order wavelength = 515.6165 \AA).
%is thought to be of atmospheric or solar origin because 
It, like the Earth's airglow emission lines (marked with
crossed circles), appear much brighter in the day+night 
spectrum than in the night-only spectrum.}
\label{fig:spectrum}
\end{figure}

Figure~\ref{fig:spectrum} displays the satellite-night and
day$+$night LiF 1A spectra for wavelengths between 1027 and 1045 \AA.
The spectra reveal \oxysix\ emission in the 1032 and 1038~\AA\ lines 
%(rest wavelengths: 1031.9261 and 1037.6167 \AA) 
(rest wavelengths: 1031.93 and 1037.62 \AA) 
and \cartwostar\ emission at 1037~\AA\ 
%(rest wavelength: 1037.0182 \AA).
(rest wavelength: 1037.02 \AA).
%in both the satellite night portion of the data and the full dataset.
An additional emission feature appears in the day+night
spectrum at 1031 \AA, but does not appear in the night-only spectrum.
It also appears in other long-duration \fuse\ daytime spectra but not
in their nighttime counterparts.  We suspect that the 1031~\AA\
feature is the second order diffraction peak associated
with atmospheric \heone, whose rest wavelength is 
%515.6165~\AA.  
515.62~\AA.  
Other \heone\ second order lines in our bandpass include
$2 \times 522.21$~\AA\ and $2 \times 537.03$~\AA, both of which 
appear in our daytime spectrum but not our nighttime spectrum.
Along with the emission lines is also a continuum, which arises from 
radioactive 
decay within the detector, high energy particles, and 
scattered terrestrial
airglow photons.  The last of these is strongest 
during the daytime portion of the orbit.

We used two methods to measure the strengths of the interstellar
emission features.  With the first method, which parallels that of 
\citet{sheltonetal01}, we fit a second order polynomial to the continuum 
of the counts spectrum. (Note that ``counts'' are called ``weights'' 
in CALFUSE version 3.1 output files.)  We then subtracted the continuum 
from the spectrum and searched for wavelength regions with large 
numbers of counts.
Due to continuum subtraction and statistical variation,
a few of the pixels within weak 
emission features  have negative numbers of counts
and some pixels outside of the emission features  have
positive numbers of counts.   
Although this noise complicated our effort to determine the boundaries
of emission features,
%In order to ensure that such statistical variation did not
%bias the measurements toward underestimating or overestimating the widths
%of the emission features, 
we set the boundaries of the emission features so as
to exclude neighboring regions exhibiting near random fluctuations.
%as the farthest high countrate pixels before regions 
%of average countrate.
We then summed the residual counts in each emission feature, 
divided by the 
instrumental effective area, and incorporated geometric 
factors in order to 
convert the measurements to units of intensity.
% Text for fits from ~ July 27, 2005:
Each measurement's statistical uncertainty was calculated 
from the square root of the number of counts in the original spectra
between the upper and lower wavelength range of the feature.

In the day$+$night spectrum, the \oxysix\ 1032~\AA\ emission line and the
daytime 1031 \AA\ feature are too close together to be separated with
this algorithm.    Therefore, when we analyzed the 1032~\AA\ 
emission line in 
the
day$+$night data, we used the wavelength range found from the 
nighttime data
(1031.64 \AA\ to 1032.45 \AA).
As can be seen in Figure~\ref{fig:spectrum}, the lower end of this
range overlaps very little with the \heone\ feature.   Furthermore,
the measured intensity in the day$+$night 1032 \AA\ \oxysix\ feature is 
less than that of the night time feature, leading us to believe
that little or no \heone\ emission was attributed to the day$+$night 
1032 \AA\ feature.
Table~\ref{table:intensityresults}
lists the measured intensities and 1 sigma 
statistical uncertainties of the
\oxysix\ 1032 and 1038 \AA\ lines, the \cartwostar\ line, and the
anomalous 1031 \AA\ feature.  
For presentation purposes, we round 
the intensities as well as the Local Bubble intensity in 
\S\ref{subsect:oviintensity} to 10s of 
photons sec $^{-1}$ cm$^{-2}$ sr$^{-1}$.
We use the unrounded values when calculating resulting quantities.
Given the uncertainties in aperture size and flux 
calibrations, additional systematic uncertainties of $14\%$ are expected
\citep{sheltonetal01}.  
Placement of the continuum is another source of uncertainty.  
Since the continuum was independently determined during each
analysis, the variation within our set of measurements 
provides an estimate of the size of this uncertainty.  The standard
deviation among the 4 measurements of the 1032 \AA\ emission line
(Table~\ref{table:intensityresults}) is 
%unrounded: 218 photons s$^{-1}$ cm$^{-2}$ sr$^{-1}$ 
220 photons s$^{-1}$ cm$^{-2}$ sr$^{-1}$ 
while the standard deviation for the 1038 \AA\ line is 
% unrounded, actually, 474.54, quoted as 
% 475 photons s$^{-1}$ cm$^{-2}$ sr$^{-1}$
470 photons s$^{-1}$ cm$^{-2}$ sr$^{-1}$.

In order to determine the central wavelengths of our irregular, weak
emission features, we fit each residual spectrum with a Gaussian and a 
quadratic.
Generally, the fitting routine modeled the emission line 
with the Gaussian and 
the residual background with the quadratic.   Occasionally, 
as occurred in
the \oxysix\ 1032~\AA\ fits, the Gaussian extended beyond the bulk of the 
emission line in order to include nearby weak residual 
background emission.
Thus, the reported wavelengths for the day$+$night 
\oxysix\ 1032~\AA\ Gaussian
fit and the nighttime \oxysix\ 1038~\AA\ Gaussian fit are 
higher and lower, 
respectively, than one would estimate by eye. 
See Table~\ref{table:velocityresults} for the tabulated results and
note that the corrected \fuse\ scale is accurate to only 
$\sim 10$ km sec$^{-1}$.

In the second method, which parallels that of \citet{dixonetal}, 
we fit the 
intensity 
spectrum's continuum with a straight line and its emission features with 
Gaussian shapes 
that had been convolved with a 106 km sec$^{-1}$ top-hat function 
(simulating 
the finite 
width of the LWRS aperture).  Because the \oxysix\ 1032~\AA\ 
emission line and 
the 
daytime 1031 \AA\  emission feature are so close together in 
the day$+$night 
spectrum, 
they were fit simultaneously.   The \oxysix\ 1038~\AA\ 
emission line and the 
\cartwostar\ 
emission line required simultaneous fitting for the same reason. The fits 
occasionally 
included outlying emission, which shifted the reported 
centroid wavelengths and 
increased 
the areas under the curves.  
%We determined the 1 $\sigma$ error bars for each model 
%parameter by increasing the best fit value of that parameter, 
%while re-optimizing the 
%other model parameters, until $\chi^2$ increased by 1.0 \citep{avni}. 
The $1 \sigma$ error bars for each model parameter were determined as
described in \citet{dixonetal}.
See Tables~\ref{table:intensityresults} and \ref{table:velocityresults} 
for the measured intensities and velocities.

The \oxysix\ and \cartwostar\ appear to be moving very slowly
%(i.e. within $\pm \sim 20$ km sec$^{-1}$)
relative to the Local Standard of Rest.   
However, 
the 1032 \AA\ signal is offset by about 30 km s$^{-1}$ from
the 1038 \AA\ signal, even
in the fits that were not corrupted by outlying emission,
the night-only 1032 \AA\ and the day+night 1038 \AA\ fits.
The difference may be due
to inaccuracies in the wavelength scale, being as the observed 
difference is somewhat more than twice the uncertainty in the velocity 
of each feature.   It could also be due to
variations in self-absorption optical depth as a function of wavelength.
Larger differences have been seen in other spectra, for example
the Virgo and Coma spectra of \citet{dixonetal}.
In the night-only 1032 \AA\ and the day+night 1038 \AA\ fits,
the $\sigma$ of the cosmic emission lines' Gaussian profiles
were 47 and 39~km~s$^{-1}$, respectively.
%The resulting Doppler parameters are 65 and 55~km~s$^{-1}$.
The resulting full widths at half maximum
are 111 and 92 km~s$^{-1}$, respectively, 
while the Doppler parameters are 65 and 55~km~s$^{-1}$, respectively.
The true widths of the cosmic emission lines should be slightly 
narrower than these values because the fitting method accounts 
for only the broadening due to a finite slit width and not any other 
form of instrumental broadening.

As a result of applying two analysis methods 
% to two independent datasets
to two dataset variants (the day $+$ night data and the night only data), 
we have 4 measurements for each of the \oxysix\ 1032 \AA, 
\oxysix\ 1038 \AA,
and \cartwostar\ 1037 \AA\ emission lines.  
For the most part, the $ 1 \sigma$ measurements overlap.
However, for each cosmic line, 3 of the 
4 intensity measurements are similar to each other while the 
4th measurement is an outlier.
From the 3 clustered and unrounded measurements, we calculate average 
intensities of the 
\oxysix\ 1032~\AA\ line, 1038~\AA\ line, and the \cartwostar\ line.
They are 
% unrounded 3189$\pm$448$, 1516\pm$347, and 1597$\pm$326
3190$\pm$450$, 1520\pm$350, and 1600$\pm$330
photons sec$^{-1}$ cm$^{-2}$ sr$^{-1}$, respectively.  
%%As Table~\ref{table:velocityresults} shows, 
%%the two methods found similar line centers in each particular case.
%The \oxysix\ and \cartwostar\ appear to be moving very slowly
%%(i.e. within $\pm \sim 20$ km sec$^{-1}$)
%relative to the Local Standard of Rest.   

Our \oxysix\ signals are neither especially bright nor
especially dim when compared with the mid latitude 
($35\arcdeg \leq b \leq 60\arcdeg$)
\fuse\ detections in the \citet{otte_dixon} and \citet{dixon_etal_06}
survey.   The larger of these compilations, that of 
\citet{dixon_etal_06}, lists 17 sight lines having 2 $\sigma$
and 3 $\sigma$ detections of
moderate velocity (-100 km s$^{-1}$ $\leq v \leq$ 100 km s$^{-1}$)
%2 and 3$\sigma$ detections of 
\oxysix\ 1032 \AA\ emission.
Of these 17 signals, 10 were brighter and 7 were dimmer
than our 1032 \AA\ signal.

We also examined the data for signs of other cosmic emission lines.  No 
significant 
signals were found.  
Our search for \carthree\ emission at 977.02 \AA\ in the 
satellite-night SiC 1B spectrum 
yielded an intensity $\pm 1\sigma$ of
% unrounded: 1304.7 $\pm$ 1742 photons sec$^{-1}$ cm$^{-2}$ sr$^{-1}$. 
1300 $\pm$ 1740 photons sec$^{-1}$ cm$^{-2}$ sr$^{-1}$. 
Similarly, our search for \nittwo\ emission at 916.71 \AA\ 
in the same spectrum yielded 
% unrounded:  510 $\pm$ 1252 photons sec$^{-1}$ cm$^{-2}$ sr$^{-1}$. 
510 $\pm$ 1250 photons sec$^{-1}$ cm$^{-2}$ sr$^{-1}$. 
Although neither species is observed in our dataset, 
both are observed at the 
$\geq2\sigma$ level in the on-filament observation 
(\carthree: 4700 $\pm$ 1300 and 
2600 $\pm$ 1000 photons sec$^{-1}$ cm$^{-2}$ sr$^{-1}$ 
in the nighttime SiC 1B 
and SiC 2A spectra, respectively; \nittwo:  
2300 $\pm$ 1100 photons sec$^{-1}$ cm$^{-2}$ sr$^{-1}$ 
in the night time SiC 1B 
spectrum, \citet{shelton_2003}).  Perhaps the observed 
photons came from the 
filament itself.  Though, if this is the case, then the 
\carthree\ intensity 
is remarkably bright, considering that theoretical 
predictions for evaporating 
clouds \citep{slavin}, predict an order of magnitude dimmer emission.

\subsection{Determination of the  \oxysix\ 
Intensity from the Halo}\label{subsect:oviintensity}

% The values for column density, electron density, etc. were updated
% in May, 2006.
%We have detected 4705 $\pm$ 567 photons cm$^{-2}$ s$^{-1}$ sr$^{-1}$ 
%the \oxysix\ resonance doublet along our line of sight.
Along our line of sight, the combined intensity arising from 
both members of 
the \oxysix\ doublet is
%unrounded:  4705 $\pm$ 567 photons cm$^{-2}$ s$^{-1}$ sr$^{-1}$.
4710 $\pm$ 570 photons cm$^{-2}$ s$^{-1}$ sr$^{-1}$.
The observed intensity can be apportioned between the Local Bubble 
(plus heliospheric) and `halo' components 
if we assume that the local emission is approximately constant over
small angular separations.  
The observation of an opaque filament on a nearby direction, 
$\ell=278.6^{\rm{o}},b=-45.3^{\rm{o}}$, set upper limits on the 
local \oxysix\ intensity \citep{shelton_2003}.   
For the local contribution,  we take the 
tightest 1 $\sigma$ upper limit 
%best-fit value
(due to random variation only) 
on the doublet from
the day$+$night data for the 1032 \AA\ emission line and the assumption
that the 1038 \AA\ line is half as strong as the 1032 \AA\ line.
This doublet intensity is
%A doublet intensity of 
%unrounded with symmetric error bars: $26^\pm{339}$ 
%rounded with symmetric error bars:
$30\pm{340}$ photons cm$^{-2}$ s$^{-1}$ sr$^{-1}$.
However, since the Local Bubble cannot produce negative photons,
we adjust the estimated intensity
arising within 230 $\pm$ 30~pc of the Earth (thus within 
%unrounded:  163 $\pm$ 21~pc
160 $\pm$ 20~pc of the Galactic plane) to
%unrounded with asymmetric error bars: $26^{+339}_{-26}$ 
%rounded with asymmetric error bars 
$30^{+340}_{-30}$ photons cm$^{-2}$ s$^{-1}$ sr$^{-1}$.
Thus, for our sight line, the remainder and vast majority of the 
observed intensity 
%unrounded with asymmetric LB error bars:  ($4680^{+567}_{-660}$
%                                           to $6227^{+753}_{-825}$ 
%rounded with asymmetric LB error bars: 
($4680^{+570}_{-660}$
%unrounded with symmetric LB error bars: ($4680\pm{660}$ 
photons cm$^{-2}$ s$^{-1}$ sr$^{-1}$)
is produced in the halo.

% SMS revision, with a few modifications by RLS:
Note that the filament data were later included in the 
\citet{dixon_etal_06} catalog.   To understand the \citet{dixon_etal_06}
reanalysis, the reader needs to first know that the
filament was observed in 5 sets of grouped LWRS exposures, 
taken over a 2 year period, towards 3 very similar directions 
on the sky 
(($\ell=278.58^{\rm{o}},b=-45.31^{\rm{o}}$), 
($\ell=278.59^{\rm{o}},b=-45.30^{\rm{o}}$), and 
($\ell=278.63^{\rm{o}},b=-45.31^{\rm{o}}$)).  Figure 3 of
\citet{shelton_2003} shows that all three directions point toward
a dense knot in an infrared-bright filament.
\citet{shelton_2003} separately plotted the spectra for
each of the three directions, found no important systematic 
differences between the spectra, then
added the spectra and searched for emission in the \oxysix\
1032 and 1038 \AA\ lines.   No emission was found near the
Milky Way's Local Standard of Rest (LSR) velocity in either
line, in either the data taken during the satellite night portion 
of the orbit or data taken during the
the satellite day + night portions of the orbit.   In our
shadowing analysis (previous paragraph), we have
adopted the tightest $1\sigma$ upper limit for the doublet,
excluding systematic uncertainties, from the
\citet{shelton_2003} analysis. It was derived from the day+night 
data for the \oxysix\ 1032 \AA\ region.

The later analysis by \citet{dixon_etal_06} differed from that of
\citet{shelton_2003}
because it included only data taken during orbital night, 
searched for only the
1032 \AA\ emission line, and used a different version of the pipeline, 
different
choices of pulse height cutoffs, and a different spectral
fitting algorithm.  
They noticed a high velocity feature in their
combined dataset, but determined that it was primarily associated 
with their data taken towards $\ell=278.59^{\rm{o}},b=-45.30^{\rm{o}}$.
In that direction, their
\oxysix\ 1032 \AA\ feature's velocity and intensity are
$206\pm13$ km~s$^{-1}$, and
$3.0 \pm 0.6 \times 10^3$ photons cm$^{-2}$ s$^{-1}$ sr$^{-1}$, 
respectively.
In contrast, neither of their spectra in the other directions 
revealed statistically significant
\oxysix\ 1032 \AA\ emission. From this, \citet{dixon_etal_06} 
concluded that \oxysix\ in the Magellanic Stream moving at 
$\sim200$~km~s$^{-1}$
relative to the LSR had been viewed through a previously
unknown hole in the filament and that the other observations 
viewed opaque portions of the filament.   When \citet{dixon_etal_06}
excluded the red-shifted feature from their analysis of
the combined dataset, they derived a $2\sigma$ upper limit 
on the \oxysix\
1032 \AA\ emission of $600$ photons cm$^{-2}$ s$^{-1}$ sr$^{-1}$
(see the last paragraph
of their Appendix A.2 and note that their naming convention differs
from that of \citet{shelton_2003}), which is roughly similar to the
$2\sigma$ upper limit reported by \citet{shelton_2003}. 
Given that they did not find
LSR rest frame emission,
they did not modify the conclusions
of \citet{shelton_2003} regarding the physical conditions
in the Local Bubble.

If the high velocity \oxysix\ 
reported by \citet{dixon_etal_06}
were to lie along the off-filament line
of sight analyzed in this paper, then its 1032 \AA\ emission 
would be $90\%$ as bright as
and 0.6~\AA\ longwards of the \oxysix\ 1032 \AA\ feature we observed.
As can be seen in Fig.~\ref{fig:spectrum}, 
our spectra are dim around $\lambda = 1032.64$~\AA, the
wavelength of \oxysix\ traveling at 206~km~s$^{-1}$.
%of the feature seen by \citet{dixon_etal_06}.   
Our wavelength calibration in this part of the spectrum is good to within
$\sim$0.034~\AA\ ($\sim$10~km~s$^{-1}$); 
therefore, the observed velocity of the
\oxysix\ 1032 \AA\ feature in our spectrum
is inconsistent with the reported velocity of the
\citet{dixon_etal_06} feature.   Thus, we conclude that the
high velocity, presumably extra-galactic feature seen by 
\citet{dixon_etal_06} 
in the I2050501 + I2050510 data has not appeared in the off-filament 
data and does not affect our analysis of the halo's emission.

Next we consider the effects of extinction.
Our off-filament line of sight is only mildly extincted,
but we obtain somewhat different results from different datasets.
According to the {\it{DIRBE}} and {\it{IRAS}} 
data \citep{schlegel_etal}, its
color excess is E(B-V) = 0.0217, implying that $N({\rm H~I})=
1.06\times 10^{20}{\rm cm}^{-2}$ if we use the empirical
relation $N({\rm H~I})/E(B-V)=4.93\times 10^{21}{\rm cm}^{-2}$
derived by \citet{diplas_savage}.  Thus, we expect that 
%unrounded:  24.74$\%$ 
25$\%$ 
of the photons originating beyond the obscuring material 
have been extincted
\citep{fitzpatrick}.  
However, according to the Leiden-Argentine-Bonn Survey 
\citep{kalberla_etal}, the column density of neutral hydrogen,
$N_{\rm HI}$, is $2.04 \times 10^{20}$~cm$^{-2}$ on the two nearest
lines of sight ($(l,b) = 278.69, -47.08$ and $278.80, -47.00$),
equating to a color excess of 0.0414 \citep{diplas_savage} 
and an extinction loss of 
%unrounded:  $42.00\%$ 
$42\%$ \citep{fitzpatrick}.  We 
recognize that this may slightly overestimate the amount of
hydrogen along our line of sight because some small portion of
the $0\fdg 6$ (HPBW) diameter beam could be responding to the
emission from the filament.
%If we use \citet{diplas_savage}'s relationship between
%$N_{\rm HI}$ and color excess and \citet{fitzpatrick}'s extinction
%curve we find that $42\%$ of the 1032 \AA\ photons originating
%beyond the obscuring material were extincted.
We will take $N_{\rm HI} = 2.0 \times 10^{20}$~cm$^{-2}$ as the upper
end of the possible range.   

For a lower limit to the absorption,
we consider the extreme case that a large fraction of the neutral
hydrogen might be beyond the \oxysix-emitting gas (almost all of which
must be more distant than the shadowing filament).   A limit to
that fraction is defined by an estimate for the amount of H~I
that exists within the Local Bubble.  \citet{lallement_etal_2003}
found that well in front of the distance to the filament 
(230 $\pm$ 30 pc) the equivalent widths of intervening
Na {\small{I}} D absorption in the spectra of stars in the
general vicinity of our sight line are
at least 20 m\AA\ and perhaps even as large
as 50 m\AA\  
(corresponding to $N_{\rm HI}= 0.5$ to $2.0 \times 10^{20}$~cm$^{-2}$,
respectively.)  Thus, there must be some extinction between our location
and
the \oxysix\ ions.  Allowing for small scale variations in ISM 
column densities and for possible revisions in the distance estimates,
we take $N_{\rm HI}= 0.5 \times 10^{20}$~cm$^{-2}$ as our lower
limit.   Also, we take $N_{\rm HI} = 1.0 \times 10^{20}$~cm$^{-2}$ 
as our nominal value.   
For $N_{\rm HI}= 1.0 \times 10^{20}$~cm$^{-2}$, the intrinsic intensity,
$I_{\rm OVI}$, is 
%rounded, for asymmetric LB error bars:
$6110^{+740}_{-860}$ 
%rounded, for symmetric LB error bars:
%$6110\pm{860}$ photons cm$^{-2}$ s$^{-1}$ sr$^{-1}$
%rounded, for asymmetric LB error bars: 
($11.8^{+1.4}_{-1.7} 
%rounded, for symmetric LB error bars: ($11.8\pm{1.7} 
\times 10^{-8}$ erg cm$^{-2}$ s$^{-1}$ sr$^{-1}$).
For the extreme range of possible extincting column densities, 
$N_{\rm HI}= 0.5$ to $2.0 \times 10^{20}$~cm$^{-2}$,
the intrinsic intensity, $I_{\rm OVI}$, is
%old:  $4680^{+567}_{-660}$
%unrounded for asymmetric LB error bars:  $5345^{+648}_{-754}$ 
%rounded for asymmetric LB error bars: 
$5350^{+650}_{-750}$ 
%rounded for symmetric LB error bars: $5350\pm{750}$ 
photons cm$^{-2}$ s$^{-1}$ sr$^{-1}$
%old:  ($9.0^{+1.1}_{-1.3}
%unrounded for asymmetric LB error bars: 
%($10.3^{+1.2}_{-1.45} \times 10^{-8}$ 
%rounded for asymmetric LB error bars: 
($10.3^{+1.2}_{-1.5} \times 10^{-8}$ 
%rounded for symmetric LB error bars: ($10.3\pm{1.5} \times 10^{-8}$ 
erg cm$^{-2}$ s$^{-1}$ sr$^{-1}$) to 
%unrounded for asymmetric LB error bars:  $7964^{+965}_{-1124}$ 
%rounded for asymmetric LB error bars: 
$7960^{+970}_{-1120}$ 
%rounded for symmetric LB error bars: $7960\pm{1120}$ 
photons cm$^{-2}$ s$^{-1}$ sr$^{-1}$
%rounded for asymmetric LB error bars: 
($15.3^{+1.9}_{-2.2} 
%rounded for symmetric LB error bars: ($15.3\pm{2.2} 
\times 10^{-8}$ erg cm$^{-2}$ s$^{-1}$ sr$^{-1}$).
% July 26, 2006: updated the de-absorbed intensity to use NH = 2.0E20
% rather than 2.04E20.

\subsection{An Estimate for the Halo's \oxysix\ 
Column Density}\label{NOVI_estimate}

In order to determine a characteristic electron density 
for the hot gas, we must 
know the
halo's \oxysix\ column density ($N_{\rm OVI}$).   
To estimate $N_{\rm OVI}$ 
along our line of sight, we have at our disposal only 
determinations made in
other directions about $15\arcdeg$ away \citep{wakker_etal}.
We make use of 4 such measurements and evaluate
their average column density.
This average equals $2.34 \times 10^{14}$~cm$^{-2}$ towards (1)
NGC~1705, (2) Fairall~9, (3) several targets within the
LMC, which we treat as a single data point, and (4) several targets
within the SMC, which we
also treat as a single data point \citep{savage-etal-2003}.  
While this mean 
value represents
our best estimate for our direction, we recognize that 
\oxysix\ column densities
vary markedly over angular separations of $15\arcdeg$, and 
thus our adopted 
value could
deviate substantially from the true value.

We now estimate the size of the error in our determination.  By examining 
Figure~11 of
\citet{savage-etal-2003}, we can better estimate the expected 
deviation from the 
parent 
population than simply evaluating the internal dispersion
of the nearest 4 lines of sight considered here.  
The average difference
in column density of individual sight lines in pairs with $\sim 15\arcdeg$
separations in the data set of \citet{savage-etal-2003}
is $35\%$ of the mean of each pair.   Thus, on average, each determination
differs from the mean of the two by $17.5\%$.
The root-mean-square ($rms$) deviations should be
$\sqrt{\pi/2}$ times this value if the distribution is Gaussian, 
and individual
samples drawn from the parent population have dispersions that 
are a factor of 
$\sqrt{2}$ larger,
leading to a fundamental $rms$ deviation of $31\%$ from what we 
could consider 
to be a ``true'' 
overall mean value over some arbitrarily large sector of the sky.   
% For the logic of this and the /sqrt(2) below, see Ed's e-mails, 
% April 2006
The mean of 4 randomly drawn samples from the parent population 
should deviate
from the overall mean by only half as much ($15.5\%$).  Thus, 
our estimate for 
$N_{\rm OVI}$ along our line of sight based on the other 4 measurements 
could be
in error by an amount equal to the error in the mean ($15.5\%$) 
combined in 
quadrature with
the intrinsic fluctuations in the individual lines of sight ($31\%$),
giving a total uncertainty of $34.7\%$.  Thus, we estimate the \oxysix\
column density and 1 $\sigma$ error
for our line of sight should equal 
$2.34 \pm 0.81 \times 10^{14}$~cm$^{-2}$.

From this value, we must subtract the Local Bubble's column density, which
we estimate from its \oxysix\ volume
density and radius.
Three of the white dwarfs in the \citet{savage_lehner} survey 
of nearby stars are within 30$^{\rm{o}}$
of our pointing direction.    The volume densities on these
short and nearby lines of sight range from 
$2.01$ to $10.6 \times 10^{-8}$~\oxysix~ions~cm$^{-3}$.
From Figure~6 in \citet{lallement_etal_2003}, we 
take the radius of the Local Bubble in our direction to
be 80 to 140~pc.   From the minimum radius and \oxysix\ volume
density, we estimate the minimum Local Bubble \oxysix\ column density
to be $5.0 \times 10^{12}$~cm$^{-2}$, and from the maximum 
radius and \oxysix\ volume density, we estimate it to be
$4.58 \times 10^{13}$~cm$^{-2}$.  Conservatively adopting these
values as the Local Bubble's average column density $\pm$ 1 $\sigma$ 
and subtracting them from the column density toward extragalactic
targets yields the halo column density, 
$N_{\rm OVI} = 2.09 \pm 0.84 \times 10^{14}$~cm$^{-2 }$

\subsection{An Estimate for the Halo's 
X-Ray Brightness}\label{subsect:xraybrightness}

We now use results from the ROSAT All Sky Survey to 
find X-ray intensities that apply to our two lines of sight.
We extract the 1/4~keV X-ray count rate
from the survey data for the R1 and
R2 bands ($\sim110$ to 284~eV and $\sim140$ to 284~eV, 
respectively).   
A 0.4$^{\rm{o}}$ radius disk centered on our direction 
$l=278.7^{\rm{o}},b=-47.1^{\rm{o}}$, has a combined \rosat\ R1 $+$ R2 
countrate of 
% 0.1 degree cone:  1307$\pm$192$\times 
%10^{-6}$counts s$^{-1}$arcmin$^{-2}$ 
%unrounded: 
1322 $\pm$ 55 $\times 10^{-6}$ 
%rounded: 1320 $\pm$ 60 $\times 10^{-6}$ counts s$^{-1}$ arcmin$^{-2}$ 
(\citet{snowden_etal_97}, retrieved with the X-Ray Background Tool: 
http://heasarc.gsfc.nasa.gov/cgi-bin/Tools/xraybf/xraybg.pl). 
This countrate includes contributions from the halo, Local Bubble,
heliospheric charge exchange, and extragalactic objects.   
Assuming that the Local Bubble contribution is roughly uniform 
over small angular scales
and that the heliospheric contribution is roughly time invariant,
we take the Local Bubble plus heliospheric contributions 
to equal the countrate seen toward the 
nearby filament 
%unrounded:  
(534 $\pm$ 131 $\times 10^{-6}$ counts s$^{-1}$ 
%rounded: (530 $\pm$ 130 $\times 10^{-6}$ counts s$^{-1}$ 
arcmin$^{-2}$ for a 0.1$^{\rm{o}}$ disk centered on 
$l=278.6^{\rm{o}},b=-45.3^{\rm{o}}$).
In order to estimate the extragalactic contribution, we draw upon the
parameterizations of Miyaji et al. (1998).  
They found that $I(E) = 10.0 \pm 0.5 \times E^{-1.42 \pm 0.03}$
and $I(E) = 11.5 \pm 0.7 \times E^{-1.42 \pm 0.03}$
photons cm$^{-2}$ k$^{-1}$ keV$^{-1}$ sr$^{-1}$
for photon energies, $E$, between 0.1 and 10 keV for the
two fields they studied.  We take the average:
$I(E) = 10.75 \times E^{-1.42}$ 
photons cm$^{-2}$ k$^{-1}$ keV$^{-1}$ sr$^{-1}$.
%For the extragalactic contribution, we use 
%%\citet{miyaji_etal}'s 
%the intensity parameterization of \citet{miyaji_etal}:  
%%the intensity:  
%$I(E) = 10.75 \times E^{-1.42}$ 
%photons cm$^{-2}$ s$^{-1}$ keV$^{-1}$ sr$^{-1}$, where
%$E$ represents the photon energy 
%in units of keV.  
When absorbed by galactic material
%($N_{\rm{HI}} = 2.04 \times 10^{20}$~cm$^{-2}$) 
(here we take the {\it{total}} column to be as high as 
$N_{\rm{HI}} = 2.0 \times 10^{20}$~cm$^{-2}$)  and
convolved with the \rosat\ response 
function, such a spectrum yields 
% unabsorbed: 236 $\times 10^{-6}$ counts s$^{-1}$ arcmin$^{-2}$ 
%for NH = 2.04E20: 54.4 $\times 10^{-6}$ counts s$^{-1}$ arcmin$^{-2}$ 
55.6 $\times 10^{-6}$ counts s$^{-1}$ arcmin$^{-2}$ 
in the R1$+$R2 bandpass,
hereafter termed the R12 bandpass.  
Subtracting the local and extragalactic contributions from the
observed countrate along our line of sight yields a halo
countrate of 
%unrounded: 
732 $\pm$ 142 $\times 10^{-6}$ counts s$^{-1}$ arcmin$^{-2}$.
%rounded: 730 $\pm$ 140 $\times 10^{-6}$ counts s$^{-1}$ arcmin$^{-2}$.
This rate applies if emission originates below all of the observed
neutral hydrogen.   However, we believe that the halo emission originates
above an absorbing column of 0.5 to $2.0 \times 10^{20}$~cm$^{-2}$.  
In this case, 
the intrinsic countrate would be 
%for NH = 2.04E20: 4911 $\pm$ 953 $\times 10^{-6}$ 
$779 \pm 151$ to 
%unrounded: $4771 \pm 927$ counts s$^{-1}$ arcmin$^{-2}$.
$4770 \pm 930$ counts s$^{-1}$ arcmin$^{-2}$.

For the analysis that will be presented in \S~\ref{temp_dist}, 
we will also 
require the
halo's R6 + R7 band (hereafter termed the R67 band) countrate.
The filament is too optically thin at 1.5 keV for a useful shadowing
analysis.  However, \citet{henley_etal}
have performed spectral fits with the \rosat\ and {\it{XMM}} data for
the on-filament and off-filament directions.   Their analysis takes
into account the extragalactic and Local Bubble contributions,
as well as instrumental and other sources of noise.   According to David
Henley (personal communication), the halo's intrinsic intensity
in the R67 band is $\sim30$ counts s$^{-1}$ arcmin$^{-2}$.

\section{Discussion}

%{\bf{Note to co-authors:   %I'll use 
%3189 $\pm$ 448 and 1516 $\pm$ 347 photons/s/cm$^2$/sr for the 
%1032 and 1038 \AA\ lines in the following calculations.  Also, 
%I'll report extra significant figures with the expectation of rounding 
%later.}}

\subsection{The Heavy Element Abundances}\label{subsect:abundances}

% Ed's abundance paragraph:
%It should be self evident that conclusions throughout this paper will be
Many of the calculations performed in this paper are dependent,
to varying degrees, on the assumed abundance ratios of heavy elements to
hydrogen, since we reported on observations of atomic transitions of
either oxygen (in the case of \oxysix\ emission or absorption) or an
ensemble that includes many other heavy elements (the main source of
soft x-ray emission).  Beyond an application to our observations, heavy
element abundances also influence some theoretical aspects of our
subject matter, such as all processes that depend the rate of radiative
cooling of the gas.  While we recognize that the abundances can vary
with galactocentric distance  
%(Peimbert 1999; Mart\'{\i}n-Hern\'andez et al. 2002; 
%Daflon \& Cunha 2004; Esteban et al. 2005), 
\citep{peimbert,martin-hernandez,daflon_cunha,esteban_etal},
and even from one
place to the next at a given radius from the center  
%(Edvardsson et al. 1993; Rolleston, Dufton, \& Fitzsimmons 1993), 
\citep{edvardsson_etal,rolleston_etal},
we adopt the simplest
interpretation that the abundances agree with the solar values.  Even
here, however, there are choices to be made.  While the outcomes for
determinations of the solar abundances of elements heavier than 
oxygen have been
reasonably stable through the years  
%(Grevesse \& Sauval 1998), 
\citep{grevesse_sauval},
very recently there have been some substantial downward revisions for C, N
and O, based on interpretations of line strengths in the context of
detailed models of line formation in a convective atmosphere  
%(Holweger
%2001; Allende Prieto, Lambert, \& Asplund 2002; Asplund et al. 2004). 
\citep{holweger,allendeprieto_etal,asplund_etal_04}.
Of particular relevance to our work is the change in the solar abundance
of oxygen relative to hydrogen, 
which has recently declined by 0.27~dex.  While
there has been some independent support for the new oxygen abundance, 
based on abundances found for O- and B-type stars  
%(Daflon et al. 2003), 
\citep{daflon_etal},
it is
discordant with models of the sound speed inside the Sun and the depth
of its convective zone, based on the interpretations of
helioseismological data  
%(Bahcall \& Serenelli 2004; Bahcall et al. 2004; Antia \& Basu 2006) 
\citep{bahcall_serenelli,bahcall_etal,antia_basu_06}
coupled with models that incorporated more
accurate calculations of atomic opacities  
%(Badnell et al. 2005).  
\citep{badnell_etal}.
The problem might be solved with a higher abundance of Ne to compensate 
for the decreased abundances of C, N and O  
%(Antia \& Basu 2005; Drake \& Testa 2005; Cunha, Hubeny, \& Lanz 2006), 
\citep{antia_basu_05,drake_testa,cunha_etal},
but this proposal has been met with some skepticism  
%(Asplund et al. 2005).  
\citep{asplund_etal_05}.
In view of the fact the solar abundance of oxygen might still be 
questioned, in various sections of
this paper we will discuss the consequences of adopting either the old
or new values.  In cases where we make our own calculations, we favor
the new abundances, but our use of some older, relatively complex models
discussed in \S\ref{subsect:halosnr} and \S\ref{temp_dist} were based on
the old abundances.

\subsection{The Physical Properties of \oxysix-bearing Gas (Isothermal 
Case)}\label{subsect:physicalproperties}

To introduce the basic concepts of our analysis of some relevant 
physical parameters, we start with the simplest case where the hot gas 
is isothermal and at the temperature that corresponds to that where 
\oxysix\ has its maximum ion fraction when the gas is in CIE
%collisional ionization equilibrium 
($3.2 \times 10^5$~K).
The density of electrons ($n_e$) in the plasma that bears the
\oxysix\ ions can be calculated from 
the halo's \oxysix\ 
column density derived in \S\ref{NOVI_estimate}, 
intrinsic doublet intensity derived in \S\ref{subsect:oviintensity},
and temperature
($T$, assumed to be the CIE 
%collisional ionizational equilibrium 
temperature, $3.2 \times 10^5$~K, but the calculation is relatively
insensitive to temperature if $10^5$~K $> T > 10^6$~K).
We use Equation 5 in \citet{shull-slavin}:
$n_e = (4 \pi I_{\rm OVI}) / (<\sigma \nu>_e N_{\rm OVI})$,
where
$<\sigma \nu>_e$ is the electron-impact excitation rate coefficient.
%
% for the $2s$ to $2p$ transition).
If the observed emission originates 
beyond an extincting column of 
$N_{\rm HI} = 1.0 \times 10^{20}$~cm$^{-2}$, then
% rounded version, case with asymmetric LB error bars: 
$n_e = 12.5^{+5.2}_{-5.3}\times 10^{-3} $ cm$^{-3}$,
% rounded version, case with symmetric LB error bars: 
%$n_e = 12.5\pm{5.3}\times 10^{-3} $ cm$^{-3}$,
but if it lies beyond our extreme estimates,
$N_{\rm HI} = 0.5$ or $2.5 \times 10^{20}$~cm$^{-2}$, 
respectively, then 
% rounded version, case with symmetric LB error bars: 
%$n_e = 11.0\pm{4.7}$
%or $16.3\pm{6.9}\times 10^{-3} $ cm$^{-3}$,
%respectively
% rounded version, case with asymmetric LB error bars:
$n_e = 11.0^{+4.6}_{-4.7}$ or
$n_e = 16.3^{+6.8}_{-6.9} \times 10^{-3}$ cm$^{-3}$
(see Table~\ref{table:parameters}.)

Table~\ref{table:parameters} also lists the thermal pressure 
and the depth 
of the emission region.   The thermal pressure, $p_{th}$,   
is calculated from the ideal gas law.   Taking the cosmic abundance 
of the 
elements into account results in 
$p_{th}/k = 1.92 n_e T$.
%{\bf{(new version:)}}
The depth of the emitting region, $\Delta l$, is equal to the 
column density
of \oxysix\ divided by the product of the electron density
and three calculated ratios:
(H/$e$), which equals 0.833 in a fully-ionized plasma with a cosmic
composition, $({\rm O/H})_\odot$, for which we adopt a value
$4.57\times 10^{-4}$ given by \citet{asplund_etal_04}, and the peak value
of the fractional
ion concentration arising from the balance of collisional ionizations
and various recombination processes,
%$f({\rm O~VI})(T_{\rm max})$ 
$f_{\rm O~VI}(T_{\rm max})$,
calculated by \citet{nahar_pradhan}.
Our result for
$\Delta l = N_{\rm{OVI}} / 
\{ n_e \times ({\rm{H}}/$e$) \times 
({\rm O/H})_\odot \times f_{\rm O~VI}(T_{\rm max}) \}$
is nearly a factor of 2 larger than that
presented in previous papers (i.e. \citet{sheltonetal01}), 
because the oxygen abundance estimates
of \citet{asplund_etal_04} are nearly a factor of 2 smaller than those
of \citet{grevesse_anders}.
% old version:
  Lastly, we calculate the time required 
for the gas to cool, if it were to cool solely by the emission of 
\oxysix\ resonance line photons.  
%from Equation 1 of \citet{shelton_2003}:
%its cooling time would be
%$t = \frac{3/2 k T}{4 \pi (sr) I_{\rm OVI}} N_{\rm OVI} 
%\frac{n_p}{n_{\rm OVI}}$,
%where $n_p$ and $n_{\rm OVI}$ are the volume densities of particles 
%\oxysix\ ions, respectively and $(sr)$ means steradians
(See Table~\ref{table:parameters}.)  
The timescale estimates are inversely proportional 
to the oxygen abundance.
The quoted estimates are approximately
twice as large as they would have been if we had used the
\citet{grevesse_anders} oxygen to hydrogen ratio. 
Furthermore, the quoted timescales are upper limits on the true 
cooling timescales for the present \oxysix-rich gas and 
do not include cooling
of nearby, hotter gas that may, in the future, evolve 
through an \oxysix-rich
stage. 
In the following subsections, we will compare these 
simple predictions with those of more comprehensive models.

\subsection{The Extended Baseline Spectrum}\label{subsect:xrayspectrum}

In this subsection, we create a long baseline spectrum from the halo's 
\oxysix\ and soft X-ray emission.  
For our comparison between the 1/4~keV and 
\oxysix\ intensities in \S~\ref{subsect:halosnr}, 
we must convert the \rosat\ R12 countrate 
%must be converted
to units of intensity.   In order to estimate the conversion
rate factor, we take the X-ray spectrum as that of a $T = 10^6$~K,
CIE
plasma, as determined 
%calculated 
by \citet{bloch_etal} and \citet{pietz_etal}; 
\citet{kuntz_snowden} found this 
temperature 
to be the dominant temperature component in their 
analysis of the halo's soft X-ray emission (the other component is 
$T = 3\times 10^6$~K).  As a result,
the halo's intensity in $\sim110$ to 284~eV photons is
%$2.0 \pm 0.40 \times 10^{-8}$ ergs s$^{-1}$ cm$^{-2}$ sr$^{-1}$ 
%if the emission originates below the galaxy's absorbing layer
$2.2 \pm 0.4 \times 10^{-8}$ ergs s$^{-1}$ cm$^{-2}$ sr$^{-1}$
if the emission originates above $N_{\rm{HI}}= 
0.5 \times 10^{20}$~cm$^{-2}$
%and $14 \pm 3.7 \times 10^{-8}$ ergs s$^{-1}$ cm$^{-2}$ sr$^{-1}$ if
%it originates above the galaxy's absorbing layer.
and $13.4 \pm 2.6 \times 10^{-8}$ ergs s$^{-1}$ cm$^{-2}$ sr$^{-1}$
if it originates above $N_{\rm{HI}}= 2.0 \times 10^{20}$~cm$^{-2}$.
The intensity of the entire 1/4~keV band amounts to only 
%$2/9$th 
21$\%$ of the
intensity in the \oxysix\ doublet
%($9.0^{+1.1}_{-1.3} \times 10^{-8}$ ergs s$^{-1}$ cm$^{-2}$ sr$^{-1}$)
if the halo emission originates above extincting material having
$N_{\rm{HI}} = 0.5 \times 10^{20}$~cm$^{-2}$ and 
%$8/9$th 
87$\%$ of the \oxysix\ doublet intensity if the halo emission originates 
above $N_{\rm{HI}} = 2.0 \times 10^{20}$~cm$^{-2}$.   
As an aside, it is of interest to compare these numbers with
the local (Local Bubble $+$ heliospheric) 
X-ray to \oxysix\ ratio, which is at least 1.4 for
the on-filament line of sight.  
%The 1$\sigma$ lower limit on the Local Bubble's X-ray to \oxysix\
%intensity ratio for the on-filament sight line is 9/5ths.
Thus, the local region preferentially sheds energy via the 
soft X-ray emission lines while the halo preferentially sheds energy
via the \oxysix\ emission lines.
We note that interpreting the relationship between the \oxysix\ and
X-ray emission rests on the assumption that the same regions of
space are being sampled. The OVI and X-ray data are from the same
directions, but have differing fields of view and differing optical depths
due to the dependence of optical depth on photon energy.
As a result, the assumption is not strictly
correct. The extent to which this is an issue depends in part
on the small scale structure in the ISM, but is anticipated to
affect our results less than the other simplifications involved
in our modeling.

Although a collisional ionizational equilibrium spectrum at $T = 10^6$~K 
has been found to fit the soft X-ray data, it underpredicts the \oxysix\
intensity (even when we assume that the halo emission has been 
absorbed by 
$N_{\rm{HI}} = 2.0 \times 10^{20}$~cm$^{-2}$.)
As the model's temperature is lowered, its
soft X-ray emission decreases more rapidly than its \oxysix\ emission.
Therefore, it is possible to find a single temperature 
spectrum that matches
both the halo's \oxysix\ and the 1/4 keV intensities.   
However, the assumed
temperature is significantly less than $10^6$~K, and so does not produce 
the soft X-ray band ratios observed across much of
the high latitude sky.
Therefore, we will move on to more complex models which are inspired
by physically conceivable events.

\subsection{The Expected Outcome for a Halo Supernova 
Remnant}\label{subsect:halosnr}  

We now examine the expected consequences for a line of 
sight in the halo that 
is dominated by the effects of a single SN event.  
Our rationale is that in
most explanations for the hot halo gas, the gas had been 
heated suddenly by an energetic event in the past.   
Irrespective of whether the
%In some conceptions, the
energetic events occurred above the Galactic disk (i.e. extraplanar
supernova explosions and collisions between infalling clouds and the
Milky Way) or
%.  In other conceptions, the energizing events occurred 
in the disk
%, but 
and the hot gas later rose into the halo,
%.  In either case, 
we expect the 
plasma to evolve from a recently shock heated and underionized state to
a tepid and overionized state; and we expect the 
observationally determined
\oxysix\ to soft X-ray ratio to be a useful diagnostic.   In 
the following halo SNR simulations, 
%models,
we found that the ratio of \oxysix\ to 1/4 keV emission rises
almost monotonically throughout the remnant's life,
making it a diagnostic of the remnant's maturity.
% (see below).

 We calculated the 
ratio of \oxysix\ to \rosat\ 1/4~keV intensities 
%was obtained 
from a simulated, extraplanar 
supernova remnant.    
%In 
For the simulation, we set the ambient density, explosion 
energy, and ambient nonthermal pressure 
%were set 
to 0.01 atoms cm$^{-3}$, 
$0.5 \times 10^{51}$~ergs, and 1800 K cm$^{-3}$, respectively.  
As in \citet{shelton_1999}, the
simulation included thermal conduction and non-CIE
%non collisional ionizational equilibrium 
radiative cooling.   The ionization and recombination rates 
were calculated using the \citet{gaetz_edgar_chevalier} tables and the 
spectra were calculated from the plasma's non equilibrium ionization 
level populations and the \citet{raymond_smith} algorithm.
We used the \citet{grevesse_anders} abundance tables, in which
the oxygen to hydrogen ratio is approximately a factor of 2 greater
then that found by \citet{asplund_etal_04}.  
The predicted \oxysix\ intensity roughly scales with the
adopted oxygen abundance, while the 1/4 keV
soft X-ray spectrum is mostly unaffected.

\begin{figure}
%\plotone{../figures/SNRratio.ps}
\epsscale{0.5}\plotone{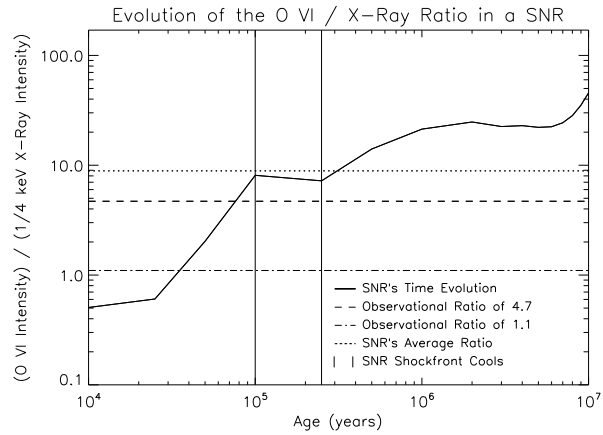}
  \caption{  
	The range of observationally derived 
	\oxysix\ to 1/4 keV intensity ratios 
%	(blue dot-dash line to red dashed line)
	(dot-dash line to dashed line)
	are compared with the predicted ratio from an evolving
	supernova remnant (thick solid line) and the remnant's average
%	ratio (black dotted line).   
	ratio (dotted line).   
	The SNR's \oxysix\ to soft X-ray intensity ratio reaches
%	1.1 (blue dot-dash line) when the remnant is about 40,000 years
	1.1 (dot-dash line) when the remnant is about 40,000 years
	old and reaches 
%	4.7 (red dashed line) when the remnant is about 70,000 years 
	4.7 (dashed line) when the remnant is about 70,000 years 
	old.
	The observationally determined ratio (1.1 to 4.7) is less
	than the remnant's lifetime
	averaged intensity weighted \oxysix\ to 1/4 keV ratio 
%	(8.9, black dotted line).    Also plotted on the graph is
	(8.9, dotted line).    Also plotted on the graph is
	the time period when the remnant evolves
	from the adiabatic (Sedov) phase to the radiative phase.
	This phase is bracketed by vertical lines on the plot.
%
%	The observationally derived ratio of 
%	\oxysix\ intensity to 1/4 keV X-ray intensity (dashed lines)
%	are compared with the predicted ratio from an evolving
%	supernova remnant (thick solid line).   
%	The observational average ratio 
%	%(4.4, assuming that the 
%	%emission originates below the halo's extincting layer) 
%	(4.7, assuming that the emitting material is extincted
%	by a column with $N_{\rm{HI}}  = 0.5 \times 10^{20}$~cm$^{-2}$)
%	crosses the SNR trajectory 
%	when the SNR is about 70,000 years old.  If, instead,
%	the emission originates above the halo's  extincting
%	layer ($N_{\rm{HI}}  = 2.0 \times 10^{20}$~cm$^{-2}$), 
%	then the ratio becomes 1.1, which crosses the
%	SNR trajectory at about 40,000 years.
%	Also marked on the graph are 
%	the time period when the remnant evolves
%	from the adiabatic (Sedov) phase to the radiative phase
%	(bracketed by vertical lines) and
%	the remnant's lifetime
%	average \oxysix\ to 1/4 keV ratio, weighted by intensity
%	(black dotted line).
	}
\label{fig:SNRratio}
\end{figure}

Here, we describe the plasma's evolution and the consequent \oxysix\ to
X-ray ratio's evolution.   When the halo supernova remnant is very young
%During its earliest years 
(age $<$ 10,000 yrs), its
%the remnant's 
most emissive portion 
is the hot, dense gas immediately behind the shock front.
Due to the rapid ionizations,
the recently shock-heated gas in this zone 
%has undergone rapid ionization, such that the 
contains \oxysix\ and higher ions.
% are present.
Although this plasma emits both \oxysix\ resonance line and soft X-ray 
photons,
% (see Figure~\ref{fig:SNRratio}), 
its temperature ($T > 10^7$~K) is far too high for optimum
\oxysix\ resonance line emission.
% ($T \sim 3 \times 10^5$~K).
%{\bf{(Note that the \oxysix/oxygen ratio has a little spike in
%the region just behind the shock and that as time goes by, there
%continues to be \oxysix\ in that location.)}}
As a result, the ratio of \oxysix\ to soft X-ray intensities is less
than the observed ratio (see Figure~\ref{fig:SNRratio}).
As time passes, the shock front slows, 
thus heating the gas in and just behind the shock to lesser temperatures
and leading to a slight increase in the \oxysix\ to X-ray ratio.

Before the remnant is $\sim50,000$ years old, 
%Since the earliest times,
both the hot dense gas near the shock front and the hot rarefied gas in 
%the remnant 
its interior are
%have been 
drastically out of CIE.
%However, 
When the remnant is between 50,000 and 100,000~years old, 
most of its plasma comes
into CIE,
%collisional ionizational equilibrium, 
but not in the usual manner.
Rather than maturing sufficiently for its ionization levels to come
into equilibrium with the gas temperature, the gas temperature
drops sufficiently as to match the ionization levels (see Figure~4 of
\citet{shelton_1999}).  At this time, the most important gas, 
that immediately 
behind the shock, is close to, but not yet in CIE.
%collisional ionizational equilibrium.
%  
During or just before this era, the simulation's \oxysix\ to
1/4~keV ratio  
crosses that of the observed ratio
%(4.4, assuming that the intrinsic emission was not absorbed, to
%(4.7, assuming the intrinsic emission was absorbed by a column of
%only $N_{\rm{HI}} = 0.5 \times 10^{20}$~cm$^{-2}$, to
(1.1, assuming that the intrinsic emission had been 
absorbed by a $2.0 \times 
10^{20}$~cm$^{-2}$ column, 
to 4.7, assuming that the intrinsic emission had been absorbed by 
a $0.5 \times 10^{20}$~cm$^{-2}$ column).
These observationally determined ratios are marked on 
Figure~\ref{fig:SNRratio}
by horizontal dashed and dot-dashed lines which
cross the
SNR ratio curve when the remnant is $\sim$40,000 and 
$\sim$70,000~years old,
respectively.
%some of the intrinsic emission has been absorbed).

As the remnant continues to cool by adiabatic
expansion and radiative cooling, its atoms begin to recombine.  
When the remnant is between 100,000 and 250,000~years, 
its shock front becomes too cool to produce much \oxysix\ 
or soft X-ray emission.   Thus the shock front no longer 
outshines the bubble's interior.   
The remnant's \oxysix\ to \rosat\ 1/4~keV ratio, now primarily 
due to the remnant's ``overionized'' interior, rises.
Henceforth, recombinations from
\oxyeight\ to \oxyseven\ to \oxysix\ provide the remnant's center with
\oxysix\ ions while reducing its supply of 
higher, more X-ray emissive, ions.
%
%The 
In time, the temperature drops from $T \sim 10^6$~K
down to $T \sim 10^5$~K,
causing the soft X-ray emission function to plummet
(although the \oxysix\ resonance line emission function remains nearly
constant.)    
As a result, the \oxysix\ versus 1/4~keV ratio continues to rise.
In its final million years, the remnant is 
tepid ($T =$ several $\times 10^4$~K), contains some \oxysix\ ions, but 
few higher ions.   Thus it produces an enormous 
\oxysix\ to 1/4~keV ratio, 
as shown in the plot.  

%For comparison with the observations, 
%two thick, horizontal dashed lines 
%mark the observationally determined \oxysix\ to 1/4~keV ratio 
%%(4.4 to 1.1).
%(1.1 to 4.7).
%The dashed lines intersect 
%the SNR ratio when the remnant is about 
%40,000 and 70,000
%years old, respectively.  
%Younger remnants contain too much hot, high pressure, recently shocked 
%gas, making them overly bright in soft X-rays.   
%Older remnants, whose hot 
%gas was shocked in the distant past, contain extensive 
%zones of \oxysix-rich 
%gas surrounding dwindling supplies of X-ray emitting gas.   
%Their \oxysix\ 
%to X-ray ratios are much larger than the observationally 
%determined ratio.  
We observe similar trends in simulated supernova remnants having greater 
ambient densities, ambient nonthermal pressures, and/or 
explosion energies, 
though with some variation in the 
age when the model matches the 
observational ratio.
Assuming that the observed gas can be compared to that in an 
undisturbed, extraplanar supernova remnant bubble,
the time since heating and the  lifetime of this gas are on the order 
of $10^4 - 10^5$ and $10^7$~years, respectively.  
The cooling timescale exceeds that calculated directly from the
\oxysix\ data (Table~\ref{table:parameters}) 
because the SNR contains a reservoir of hotter,
more highly ionized gas that will eventually 
transition through the \oxysix\ level.
%Irrespective of the specific source of the halo's hot gas along our 
%line of sight, whether supernova remnants, fountains, 
%or infalling clouds,
Assuming that the other possible sources of hot gas behave similarly to
simulated halo SNRs,
we suggest that the resulting structure is middle aged. 
%It is neither very
%recently shocked gas, nor an ancient pool of gas from 
%one or more very old 
%events because very young or very old gas would probably not be able 
%to produce the observed \oxysix\ to soft X-ray ratios.

\subsection{The Volume Distribution Function of 
Temperatures}\label{temp_dist}

\subsubsection{The Basic Assumptions}\label{subsubsect:assumptions}

We move on to a more generalized picture and propose 
that the hot gas in the 
halo of our Galaxy is a heterogeneous mixture of regions 
having plasmas at 
different temperatures, created possibly by many SN events 
whose influences on 
the halo medium have merged together.  A convenient 
characterization of the 
temperature distribution function over volume can take 
on the form of a power-law, 
one that extends from $10^5\,$K (above which appreciable amounts of 
\oxysix\ are expected in CIE) to a sharp cutoff at some high temperature 
$T_{\rm{cut}}$.  Within the paradigm that the hot gas is 
created by supernova 
shock waves that are produced in the halo or that escape 
from the disk, the 
origin of $T_{\rm{cut}}$ can be interpreted to arise from 
either a generalized 
limit on the supernova shock velocities or, alternatively, 
from our recognition 
that gases with temperatures that are too high may escape 
very rapidly in the 
form of a very low density, high velocity wind, which is 
difficult to detect.

Since all of the observable effects of the hot gas 
represent line integrals 
of various physical quantities through the Galactic halo, there is some 
benefit in our starting with a formulation that describes how the many 
differential length segments $dl$ within the population of discrete, 
homogeneous gas regions are distributed over temperature.  We do this by 
specifying a transformation between $dl$ and $d\ln T$,
\begin{eqnarray}\label{dl}
dl&=&BT^\beta d\ln T~{\rm for}~T<T_{\rm cut}\cr
&=&0~{\rm for}~T>T_{\rm cut}~,
\end{eqnarray}
where $B$ is a distance scale factor.  In a broad, 
statistical sense, this 
distribution function describes how temperatures are 
weighted according to 
volume fractions, but it says nothing about the internal 
electron densities 
$n_e$ within the length segments.

In the development of our interpretation, we impose a 
simplifying constraint 
that the Galactic halo is approximately isobaric.  We use this universal 
pressure constraint to tie the electron density to temperature and
thermal pressure, i.e., 
%$n_e=p/(1.91kT)$.  
$n_e=p_{th}/(1.92kT)$.  
While we must accept the reality that some pressure 
variations can exist from one location to another, we can assume that the 
magnitude of such variations are small compared to the vastly different 
temperatures that we adopt in our model.  Our
representative pressure 
$p_{th}$ is a free parameter that we will determine from the ratio of 
$I_{\rm OVI}/N_{\rm OVI}$ after we have solved for the power-law 
coefficient $\beta$ and temperature limit $T_{\rm cut}$.

\subsubsection{How the Observations Relate to the 
Model}\label{subsubsect:relate}

The column density of \oxysix\ is given by the expression
\begin{equation}\label{NOVI_dl}
N_{\rm OVI}=\left( {{\rm O}\over {\rm H}}\right)_\odot 
\left( {{\rm H}\over e}\right)~\int f_{\rm OVI}(T) n_e dl
\end{equation}
where $(\rm O/H)_\odot$, $({\rm H}/e)$, and $f_{\rm OVI}(T)$ were 
first used in \S\ref{subsect:physicalproperties}.
%A fully-ionized plasma with a cosmic composition
%should have $({\rm H}/e)=0.833$.  
We can rewrite Eq.~\ref{NOVI_dl} in
terms of an integral over $\ln T$ of the temperature distribution we
adopted,
\begin{equation}\label{NOVI_dlnT} 
N_{\rm OVI}=\left( {{\rm O}\over {\rm H}}\right)_\odot 
\left( {{\rm H}\over e}\right)~{p_{th}B\over 1.92k}
\int_{\ln 10^5 \rm{K}}^{\ln T_{\rm cut}}f_{\rm OVI}(T)T^{\beta -1}d\ln T
\end{equation}
by making a substitution of the terms in Eq.~\ref{dl} for $dl$, 
using the relation $n_e=p_{th}/(1.92kT)$,
and noting that \oxysix\ is rare at temperatures below $10^5$~K.

We can develop similar equations for the expected strengths of the line
radiation from \oxysix\ or the intensities of soft X-rays.  
For the emission
from both members of the \oxysix\ doublet, we anticipate that
\begin{equation}\label{IOVI_dl}
I_{\rm OVI}=\int r_{\rm OVI}(T)n_e^2 dl~,
\end{equation}
where the emission rate coefficient $r_{\rm OVI}(T)$ per unit emission
measure is given by
\begin{equation}
\label{rOVI} 
r_{\rm OVI}(T)=3.09\times 10^{18}\left( {{\rm O}\over {\rm H}}
\right)_\odot \left( {{\rm H}\over e}\right)f_{\rm OVI}{<\sigma v>(T)
\over 4\pi}~{\rm photons~cm}^{-2}{\rm s}^{-1}{\rm sr}^{-1}({\rm cm}^{-6}
\,{\rm pc})^{-1}~.
\end{equation}
As we did in \S\ref{subsect:physicalproperties},
we adopt an analytical formulation of $<\sigma v>(T)/(4\pi)$ that was
specified by Shull \& Slavin (1994).  
Again after substituting the expression in Eq.~\ref{dl} for $dl$ and 
$p_{th}/(1.92kT)$ for $n_e$ we find that
\begin{equation}\label{IOVI_dlnT}
I_{\rm OVI}={p_{th}^2B\over (1.92k)^2}
\int_{\ln 10^5 \rm{K}}^{\ln T_{cut}} 
r_{\rm OVI}(T)T^{\beta -2}d\ln T~
\end{equation}
A similar equation can be expressed for the soft X-ray emission,
\begin{equation}\label{R12_dlnT}
R_{12}={p_{th}^2B\over (1.92k)^2}\int_{\ln 10^5 \rm{K}}^{\ln T_{cut}} 
r_{12}[N_{\rm HI},T]
T^{\beta -2}d\ln T~,
\end{equation}
where 
$r_{12}[N_{\rm HI},T]$ is the emission rate coefficient as a
function of temperature for 0.25~keV X-rays, matched to the responses of
the \rosat\ bands 1 and 2 and expressed in the units 
${\rm counts~cm}^{-2}\,{\rm s}^{-1}\,{\rm arcmin}^{-2}({\rm cm}^{-6}\,
{\rm pc})^{-1}$ (Snowden et al. 1997).\footnote{The cosmic 
abundances adopted
by Snowden et al. (1997) did not incorporate the recent downward
revisions of the solar photospheric abundances of C (Allende Prieto et
al. 2002), N (Holweger 2001) or O (Asplund et al. 2004).  The correction
to allow for these abundance changes should be rather small, since the
line emission over the energy range of interest is dominated by lines
from other, much heavier elements, such as Si, S, Mg and Fe; see, e. g.,
Kato (1976).}  This coefficient includes the effect of an
energy-dependent foreground absorption represented by $N_{\rm HI}$.

\subsubsection{Evaluation of Parameters}\label{subsubsect:parameters}

\begin{figure}
\epsscale{0.65} 
%\plotone{../figures/shelton_plts.eps}
\plotone{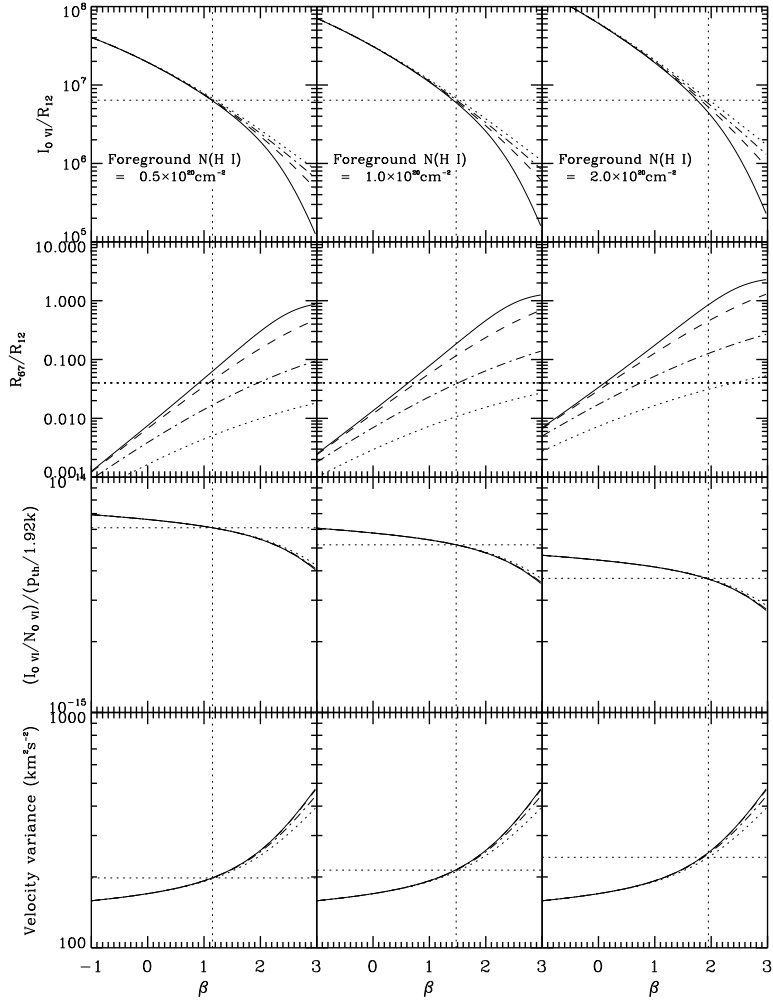}
\caption{Plots that indicate the sensitivity of various 
physical parameters on 
the values of observed quantities.  The three columns of panels represent 
solutions of the equations assuming different values for the foreground 
absorption, represented by 
$N_{\rm HI}$: ({\it left\/}) $0.5\times 10^{20}$~cm$^{-2}$, ({\it 
middle\/}) $1.0\times 10^{20}$~cm$^{-2}$, and ({\it right\/}) $2.0\times 
10^{20}{\rm cm}^{-2}$cm$^{-2}$.   In all of the plots, curves 
for different 
assumed values of $T_{\rm cut}$ are drawn in different styles: 
solid curves: 
$\log T_{\rm cut}=8.0$; dashed curves: $\log T_{\rm cut}=6.9$; 
dash-dot curves: 
$\log T_{\rm cut}=6.6$; dotted curves: $\log T_{\rm cut}=6.4$. 
{\it Top row:\/} 
The dependence of $\beta$ on the value of 
$I_{\rm OVI}/R_{12}$ in units of \oxysix\ photons arcmin$^{2}$ 
R12 counts$^{-1}$ sr$^{-1}$, 
as defined in Eq.~\protect\ref{ratio_IOVI_R12}.  
From the 
intersection of the observed value of this ratio 
(horizontal dotted line) and 
the curves, we obtain the best values of $\beta$ in each case 
(vertical dotted 
lines dropped down to the $\beta$ scales at the bottom).  
{\it Second row:\/} 
The dependence of $T_{\rm cut}$ on $R_{67}/R_{12}$ as a 
function of $\beta$, as 
defined by Eq.~\protect\ref{ratio_R67_R12}.  We used these 
curves to define 
appropriate values of $T_{\rm cut}$ in each case by locating 
the one that most 
closely intersects the horizontal dotted line representing 
the observation at 
the preferred value of $\beta$.  {\it Third row:\/} The expected ratio 
$(I_{\rm OVI}/N_{\rm OVI})/(p_{th}/1.92k)$ as a function of $\beta$, as 
expressed in 
Eq.~\protect\ref{ratio_IOVI_NOVI}.  The intersections of 
the dotted lines show 
the outcomes for the best $\beta$ values with the thermal 
pressures listed in 
Table~\protect\ref{pwr_law_parameters}.  {\it Bottom row:\/} The expected 
velocity variance as a function of $\beta$, as defined in 
Eq.~\protect\ref{vel_var}. }\label{multipanel}
\end{figure}

To evaluate the free parameters $\beta$, $T_{\rm cut}$ and $p_{th}/k$, we
compare three ratios of observable quantities with their expectations
within our formalism.  First, we may constrain the value of $\beta$ by
matching the observed ratio of $I_{\rm OVI}$ to $R_{12}$ to the
expression
\begin{equation}\label{ratio_IOVI_R12}
{I_{\rm OVI}\over R_{12}}={\int_{\ln 10^5 \rm{K}}^{\ln T_{\rm cut}}
r_{\rm OVI}(T)T^{\beta -2}d \ln T\over 
\int_{\ln 10^5 \rm{K}}^{\ln T_{\rm cut}}
r_{12}[
N_{\rm HI},T] T^{\beta -2}d \ln T}~.
\end{equation}
While the result depends on an adopted value of $T_{\rm cut}$, for
reasonably high values of this quantity the effect is small.  We find
that for $I_{\rm OVI}=4680\,{\rm photons~cm}^{-2}{\rm s}^{-1}{\rm
sr}^{-1}$ and $R_{12}=0.000733\,{\rm counts~cm}^{-2}{\rm s}^{-1}{\rm
arcmin}^{-2}$ we obtain $I_{\rm OVI}/R_{12}=6.4\times 10^6$ (in the
same units) with an uncertainty of 24\% using the errors stated in 
\S\ref{sect:observations}.
%and 
%\ref{subsect:xrayspectrum}.  
%\ref{subsect:xraybrightness}.
The first row of Table~\ref{pwr_law_parameters} 
lists the derived values of 
$\beta$ for the best and limiting values of 
foreground H~I absorption.  The top 
row of panels in Fig.~\ref{multipanel} shows how 
$I_{\rm O~VI}/R_{12}$ changes 
with $\beta$ and demonstrates that the dependence on $T_{\rm cut}$ is not 
important for $\beta < 1.5$ and is a small effect for $\beta < 2$.  

A strong response to $T_{\rm cut}$ arises from the ratio of high energy 
X-ray emission to the intensity at lower energies, 
as indicated in the second 
row of panels in Fig.~\ref{multipanel}, where we 
have made use of an equation 
that is identical in form to Eq.~\ref{ratio_IOVI_R12} 
except that it compares
the ratio of 
X-ray emission at 1.5~keV (the sum of intensities recorded in
\rosat\ Bands 6 and 7) to that at 0.25~keV (the sum 
in \rosat\ Bands 1 and 2),
\begin{equation}\label{ratio_R67_R12}
{R_{67}\over R_{12}}={\int_{\ln 10^5 \rm{K}}^{\ln T_{\rm cut}}r_{67}
[N_{\rm H~I},T]T^{\beta -2}d \ln T\over 
\int_{\ln 10^5 \rm{K}}^{\ln T_{\rm 
cut}}r_{12}
[N_{\rm H~I},T]T^{\beta -2}d \ln T}~
\end{equation}
From the observed value of $R_{67}/R_{12}=0.04$ 
%(\S\ref{subsect:xrayspectrum}),
(\S\ref{subsect:xraybrightness}),
we arrive at values of $\log T_{\rm cut}$ given in the second row of 
Table~\ref{pwr_law_parameters}.

Finally, we solve for $p_{th}/k$ by matching the observed ratio of 
$I_{\rm OVI}$ 
to $N_{\rm OVI}$ to the formula
\begin{equation}\label{ratio_IOVI_NOVI}
{I_{\rm OVI}\over N_{\rm OVI}}=
{p_{th}\over 1.92k}~{\int_{\ln 10^5 \rm{K}}^{\ln T_{\rm cut}}f_{\rm OVI}
[<\sigma v>(T)/(4\pi)]T^{\beta -2}d \ln T\over
\int_{\ln 10^5 \rm{K}}^{\ln T_{\rm cut}}f_{\rm OVI}T^{\beta -1}d \ln T}~
\end{equation}
Values of $p_{th}$, found for
$I_{\rm OVI}/N_{\rm OVI }=2.24\times 10^{-11}$ 
photons s$^{-1}$ sr$^{-1}$ with an uncertainty of about 50\%,
are listed in the third row of
Table~\ref{pwr_law_parameters}.  From the flatness of the 
curves shown in the 
third row of panels in Fig.~\ref{multipanel}, the value 
of $p_{th}$ is not 
strongly influenced by the choice of $\beta$ (and has 
virtually no dependence on 
$T_{\rm cut}$).  The dominant uncertainty in $p_{th}$ 
arises from the expected 
error in 
$I_{\rm OVI} / N_{\rm OVI}$. 

It is important to check that the computed width for thermal Doppler 
broadening of \oxysix\ in the model does not exceed the observed widths of
the actual absorption profiles observed in the Galactic halo.  
Our expectation is that
\begin{equation}\label{vel_var} 
<v^2>={\int n({\rm OVI})<v({\rm OVI})^2> dl\over \int n({\rm OVI})dl}
= {k\over 16m_{\rm p}}~{\int_{\ln 10^5 \rm{K}}^{\ln T_{\rm cut}} 
f_{\rm OVI}(T)T^\beta d\ln T\over 
\int_{\ln 10^5 \rm{K}}^{\ln T_{\rm cut}} 
f_{\rm OVI}(T)T^{\beta -1}d \ln T}~,
\end{equation}
where $m_{\rm p}$ is the mass of a nucleon.
For our derived values of $\beta$, we find that the Doppler parameter
$b=(2<v^2>)^{\onehalf}$ ranges from 19 to $23\,{\rm km~s}^{-1}$.  
These values 
are smaller than the representative observed values, which are of order
$60\,{\rm km~s}^{-1}$.  Evidently, even with our allowance for 
some \oxysix\
arising from temperatures above $10^6\,$K, the expected profile widths
are still much less than the kinematic broadening that takes place in
the halo.  As with our findings on $p_{th}/k$, the bottom 
row of panels in 
Fig.~\ref{multipanel} shows that the outcome from
Eq.~\ref{vel_var} is nearly independent of $T_{\rm cut}$.   

So far, all of our evaluations of various ratios of certain quantities
have not had to incorporate the constant $B$ that appears in
Eq.~\ref{dl}, since this term canceled out in each case.  This constant
is important, however, if we wish to relate the results from the
observations to either a total length scale along our line of sight, 
$L=\int dl$, or the values
for the electron density, $n_e(T)$.  By solving either
Eqs.~\ref{NOVI_dlnT}, \ref{IOVI_dlnT} or \ref{R12_dlnT} 
one can derive this 
constant, which varies markedly with the adopted value of $\beta$ (as it 
should).  Values of $B$ for the various combinations 
of $\beta$ and $N({\rm 
H~I})$ are listed in Table~\ref{pwr_law_parameters}.  
From an appropriate  value of $B$ we can evaluate the total length
of hot gas along our path
\begin{equation}\label{L} 
L=B\int_{\ln 10^5 \rm{K}}^{\ln T_{\rm cut}}T^\beta d\ln T=B(T_{\rm
cut}^\beta-10^{5\beta})/\beta~.
\end{equation} 
Outcomes for $\log L$ (in pc) are listed in 
Table~\ref{pwr_law_parameters}.
% addition:
If the gas above and below the plane is stratified in layers
parallel to the plane, then a representative vertical
sight line (i.e. one perpendicular to the galactic disk)
intersects less material than our sight line at an intermediate
galactic latitude of $b = -47.1\arcdeg$.
Thus, calculations for the vertical sight line would require
that the total length scale, $L$ in Eq.~\ref{L}
be foreshortened by a factor of $\sin(|b|)$.
Similarly, $N_{\rm{OVI}}$ in Eq.~\ref{NOVI_dlnT},
$I_{\rm{OVI}}$ in Eq.~\ref{IOVI_dlnT},
$R_{12}$ in Eq.~\ref{R12_dlnT},
together with 
$N_e$ and $dEM(T)$ in the upcoming Eqs.~\ref{Ne} \ref{EM}, 
would require reduction by the
same factor.

\begin{figure}
\epsscale{1.0}
%\plotone{../figures/shelton_plt2.eps}
\plotone{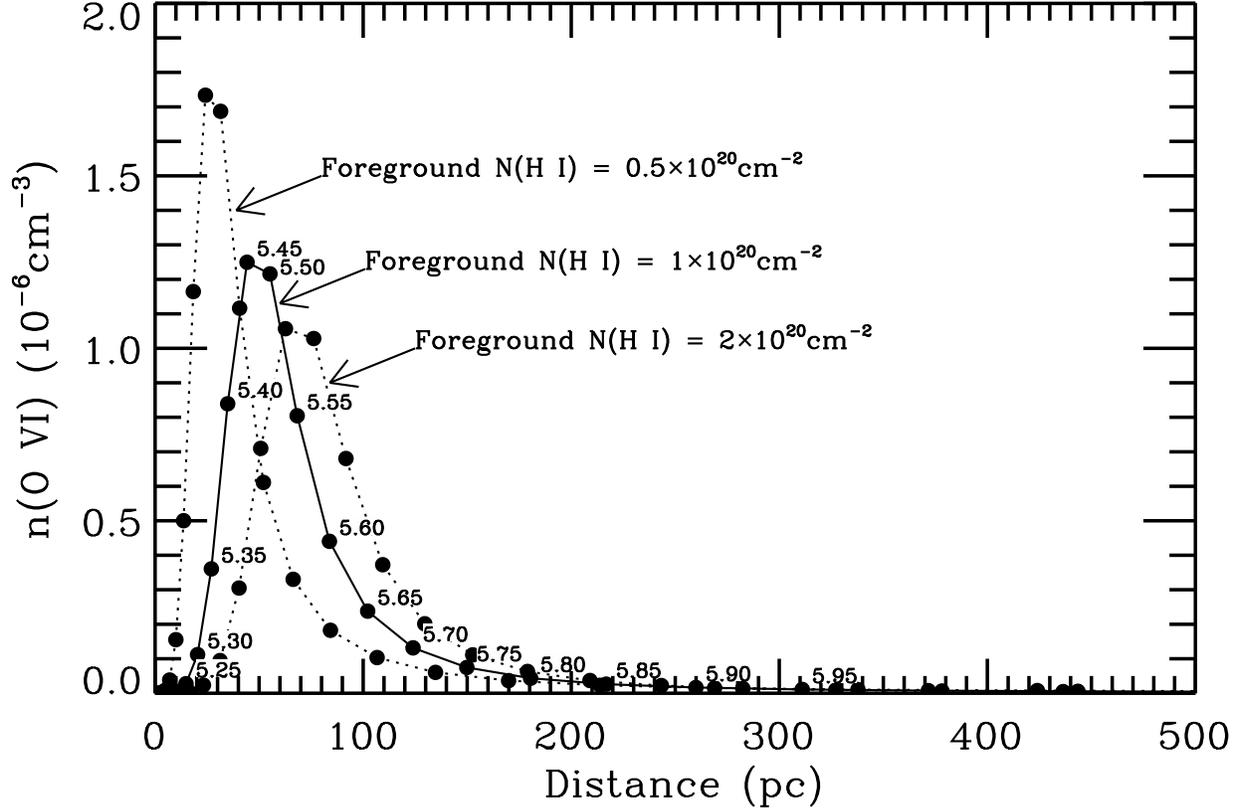}
\caption{The density of \oxysix\ vs. distance along our line of sight
for the hypothetical case where all 
of the regions are arranged in an end-to-end sequence from low to high 
temperature.  The solid curve represents our solution to the power-law 
temperature (or length-scale) relation for a foreground 
absorption equivalent to 
$N_{\rm HI}=1.0\times 10^{20}{\rm cm}^{-2}$, 
$p_{th}/k=8320\,{\rm cm}^{-3}$K, 
and $\beta=1.48$.  On this curve, points are labeled according to their 
respective values of $\log T$.  The other (dotted) curves 
represent other values 
of foreground absorption, as indicated (with $p_{th}/k$ and 
$\beta$ in each case 
set to the respective optimum values - see 
Table~\protect\ref{pwr_law_parameters}).}\label{plt2}
\end{figure}

On the one hand,  it is important to note
that $L$ represents a minimum length scale, since
parcels of gas could be scattered over a longer distance with some
unseen gas phase filling in the intervening gaps.  On the other hand,
the fraction of $L$ that has an appreciable 
concentration of \oxysix\ is small.  
Figure~\ref{plt2} illustrates what the concentration 
of \oxysix\ would look like 
if all of the gas parcels at different temperatures 
were sequenced along a line 
in order of increasing temperature.  Most of the 
\oxysix-bearing gas is confined 
within a region that is less than about 100~pc thick, 
with the remaining, much 
greater length filled with a
plasma that is too hot to contain much O VI.  This 
thickness is considerably 
less than the 2.3~kpc scale height found for \oxysix\ 
in the Galactic halo by 
Savage et al. (2003).  Evidently, our large observed ratio of 
$I_{\rm OVI}/N_{\rm OVI}$ 
shows that the gas with temperatures of around 300,000~K 
is highly clumped, even 
when one acknowledges that there can be a broad 
distribution of temperatures.  
The internal densities 
$n({\rm OVI})$
within each clump, which in our model can reach as high as 
$\sim10^{-6}~{\rm cm}^{-3}$, 
is about two orders of magnitude higher than an overall average 
$<n({\rm OVI})>=1.7\times 10^{-8}{\rm cm}^{-3}$
that was found in a \fuse\  survey of the Galactic disk 
\citep{bowen_etal}. 

The column density of electrons along our line of sight is given by
\begin{equation}\label{Ne}
N_e=B\int_{\ln 10^5 \rm{K}}^{\ln T_{\rm cut}}n_e(T)T^\beta d\ln T
={p\over 1.92k}~{B(T_{\rm cut}^{\beta -1}-T^{5(\beta-1)})\over \beta -1}~,
\end{equation}
and we list outcomes for the appropriate values of $\beta$ and 
$N_{\rm HI}$
in Table~\ref{pwr_law_parameters}. 

% one of Ed's additions, November 16, 2006
%
In \S\ref{subsect:abundances} we discussed the possibility that the
abundance of oxygen might be higher than the one that we adopted from
\citet{asplund_etal_04}.  If we were to adopt the old, higher abundance
value given by \citet{grevesse_sauval} (larger by 0.27~dex), the values
of $\beta$ listed in Table~\ref{pwr_law_parameters} would increase by
about 0.5 because the relative proportion of the plasma at temperatures
that emit \oxysix\ line radiation is reduced, when compared to the
higher temperature material that emits soft X-rays.  (Recall that the
predicted X-ray emission is not strongly influenced by changes in the
abundance of oxygen, as we stated in an earlier footnote.)  For the three
values of foreground $N$(H~I), (0.5, 1.0, $2.0\times 10^{20}\,{\rm
cm}^{-2}$), the old oxygen abundance changes $\log p_{th}/k$ by (+0.03,
+0.03, +0.05)~dex, respectively.  Values of $\log L$ change by (+0.23,
+0.09, +0.17)~dex, while $\log N_e$ changes by (+0.05, +0.01,
+0.03)~dex.

\subsubsection{The Total Radiative Loss Rate for the 
Halo}\label{subsubsect:lossrate}

	Here, we assume that the observed line of sight
provides a fair representation of the halo as a whole.
This assumption allows us to calculate the total radiative
energy loss rate associated with hot gas in the halo,
a rate that can be usefully compared with the energy 
injection rate.

To compute the radiative energy loss rate 
from the hot gas in the halo,
we must know how the differential emission 
measure $dEM=n_e^2(T)dl$ varies with
temperature.  With our reformulation of $dl$ in terms of $d\ln T$ we
find that
\begin{equation}\label{EM}
%EM(T)=\left({p_{th}\over 1.92k}\right)T^{\beta -2}d\ln T
dEM(T)=\left({p_{th}\over 1.92k}\right)^2 B T^{\beta -2}d\ln T
\end{equation}
along our line of sight.
The second to the last group of numbers in Table~\ref{pwr_law_parameters} 
shows the logarithms 
of the coefficient in front of the $T^{\beta - 2}$ term in the equation. 

\begin{figure}
\epsscale{1.0}
%\plotone{../figures/shelton_plt3.eps}
\plotone{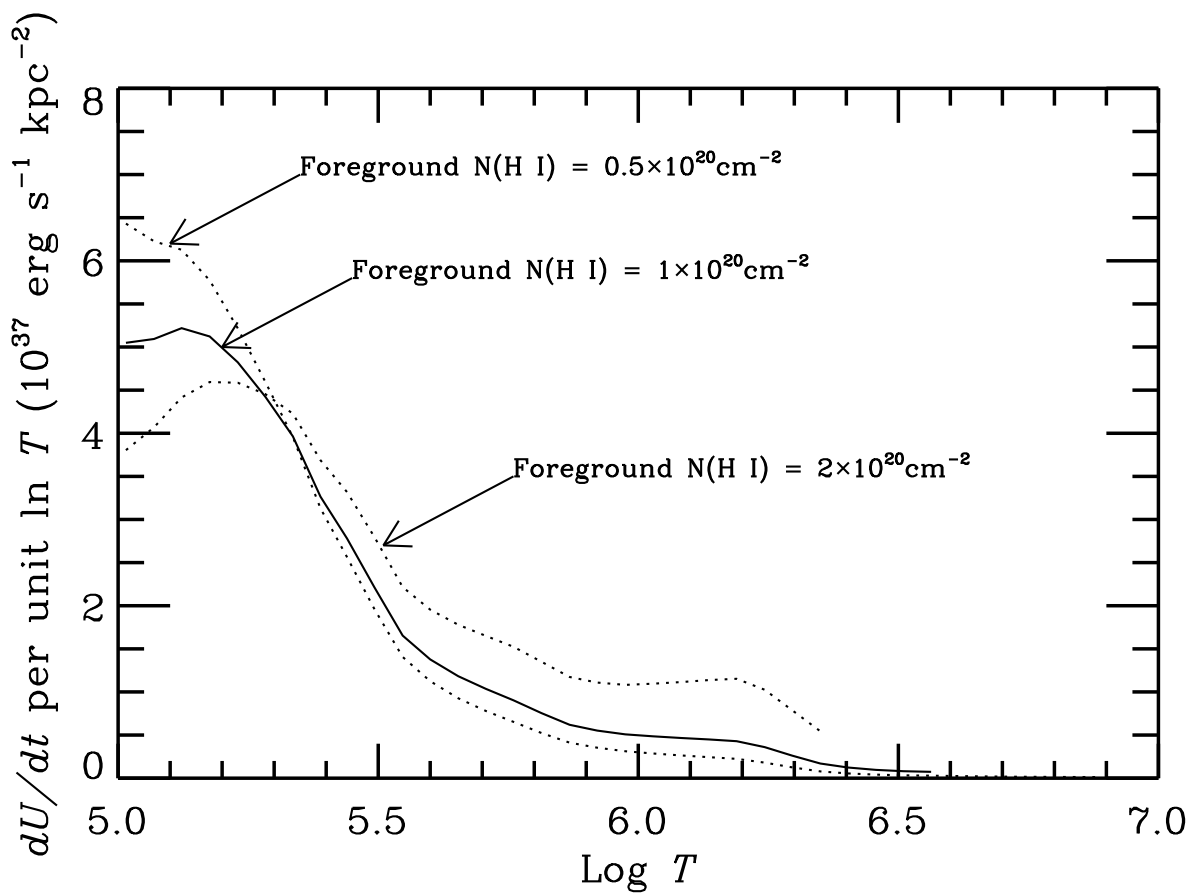}
\caption{Radiative energy loss rates per unit area, $dU/dt$, as a function
$\log T$ for the preferred values of $\beta$ that apply to the three
possible values of the foreground $N$(H~I) (see
Table~\protect\ref{pwr_law_parameters}).  The curves plotted represent
the product of the integrand and constants given in
Eq.~\protect\ref{dEdt}.}\label{plt3}
\end{figure}

With our expression for the emission measure as a function of $T$, we
are now prepared to estimate the power radiated by the gas at
temperatures above $10^5\,$K.  For a cooling function $\Lambda_N$ that
is normalized to the local product of electron and ion densities
$n_en_i$, we adopt the values listed by \citet{sutherland_dopita} that
apply to a plasma that has a solar composition and nonequilibrium ion
fractions in a regime of radiative, isochoric cooling, starting with an
initial temperature of $10^{7.5}\,$K.  If we were to suppose that our
line of sight shows a fair representation of the cooling per unit area
$dU/dt$ by hot gas on both sides of the Galactic plane in our region of
the Galaxy, we find that
\begin{equation}\label{dEdt}
dU/dt=2\sin(|b|)(0.917)
\int_{\ln 10^5\rm{K}}^{\ln T_{\rm cut}}EM(T)\Lambda_N 
d\ln T
\end{equation}
where $b$ is the Galactic latitude of our line of sight ($-47\fdg 1$),
the factor 2 accounts for both sides of the plane, and the factor 0.917
allows for the fact that our expression for $EM(T)$ is cast in terms of
$n_e^2$ whereas the normalization of the function expressed by
\citet{sutherland_dopita} is normalized to $n_en_i$.  Values of $\log
dU/dT$ are listed in the last group of numbers in
Table~\ref{pwr_law_parameters}.  Several important qualifications must
be expressed about these numbers.  
%First, we have truncated the
%calculation at a lower limit $T=10^5\,$K, but significant additional
%radiation could arise from gas at somewhat lower temperatures.  
First, we have truncated the calculation at a lower limit $T=10^5\,$K
because we have no ability to sense gas at lower temperatures.  If an
extrapolation of our power-law representation seems plausible, we should
expect to find some additional energy radiated below $10^5\,$K.  This is
demonstrated in Fig.~\ref{plt3}, where we have plotted the shape of the
integrand in Eq.~\ref{dEdt} as a function of $\log T$.  Second,
\citet{sutherland_dopita} used the solar abundances given by
\citet{anders_grevesse} instead of the more modern values of
\citet{allendeprieto_etal} that we adopted.  
% One of Ed's additions, November 16, 2006:
Thus, if the function
$\Lambda_N$ were to be recalculated using the newer abundances, we would
find a somewhat lower cooling rate, but by a factor that is less severe
than the changes in the C, N and O abundances, since other elements are
also important coolants.  If indeed the older abundances are correct,
the ripple effect from the changes in the parameters derived in
\S\ref{temp_dist} above could reduce the emission power by up to
-0.4~dex.  Aside from these possible offsets, our determination
should be free of bias, 
%The former had oxygen abundances
%that were 0.27~dex greater than the latter.  Thus, if the function
%$\Lambda_N$ were to be recalculated using the newer abundances, we would
%find a somewhat lower cooling rate.  Aside from this offset, 
%our determination should be free of bias, 
%
and indeed the values of \oxysix\ and X-ray
emission found here are fairly typical of those found along other
high latitude sight lines.  Nevertheless,
it is important to note that the distribution of hot gas in the
Galactic halo is extremely uneven; hence our single line of sight does
not give a very accurate representation of the average $EM$ on either
side of the Galactic plane.

The rate of cooling, $dU/dt$, is similar to the rate of 
energy input from supernova explosions and pre-supernova winds.
The average massive SN progenitor star releases $1.4$ to 
$2.0 \times 10^{50}$~ergs
in wind energy before it explodes, according to calculations in 
\citet{ferriere_1998} and \citet{leitherer_etal}.
At the Sun's galactocentric radius, 18.6 massive stars 
and 2.6 white dwarfs 
explode per Myr per kpc$^{2}$ cross-sectional disk area, 
though many of these 
stars explode above the disk \citep{ferriere_1998}.
If each explosion releases $10^{51}$~ergs of energy, then 
SN and pre-SN winds inject 
%$7.56$ to $7.92 \times 10^{38}$ erg kpc$^{-2}$ s$^{-1}$
$7.66$ to $8.06 \times 10^{38}$ erg kpc$^{-2}$ s$^{-1}$
into the ISM.
%The majority of the injected SN and pre-SN
%energy is later radiated away by hot halo gas;
%in our nominal case, the hot halo radiates
%$5.37^{+0.80}_{-0.69} \times 10^{38}$ erg kpc$^{-2}$ s$^{-1}$.
A comparison with the calculated energy loss rate for our nominal case
($5.37^{+0.80}_{-0.69} \times 10^{38}$ erg kpc$^{-2}$ s$^{-1}$)
shows that the majority of the injected SN and pre-SN
energy is later radiated away by hot halo gas.
Since the halo cooling rate accounts for $\sim70\%$ of
the SN and pre-SN wind energy injection rate, little remains to
power other activities, such as large scale galactic winds.    
Half of the photons emitted by the halo will travel toward the
galactic midplane and likely be absorbed, while the other half
will travel upwards.   Their absorption rate depends on the column
density of gas residing above the emitting gas, which is expected to
be less than $\sim1 \times 10^{20}$~cm$^{-20}$.   Thus, most of
the 1032 \AA\ and shorter photons will leave the system.

\section{Summary}\label{sect:summary}

% from RLS:  These numbers have been updated due to the change from
% NH = 0 to 2.04E20 to NH = 0.5 to 2.0E20.

We analyzed \fuse\ LWRS data for $\ell = 278.7^{\rm{o}}, 
b = -47.1^{\rm{o}}$,
finding an \oxysix\ doublet intensity of 
%unrounded: 4705 $\pm$ 567 
4710 $\pm$ 570 photons cm$^{-2}$ s$^{-1}$ sr$^{-1}$.   
Our pointing direction is only mildly extincted, 
so the observed intensity included contributions from the
Local Bubble and the Galactic halo.
Only $2^{\rm{o}}$ from our pointing direction,
the sky is heavily extincted by a filament residing 230 pc from the Earth.
A previous observation of that direction indicated the Local Bubble's
contribution.  The difference between the intensities observed on these
two sight lines, 
%unrounded with asymmetric LB error bars:  $4680^{+567}_{-660}$
%rounded with asymmetric LB error bars: 
$4680^{+570}_{-660}$ 
%unrounded with symmetric LB error bars: $4680\pm{660}$
%rounded with symmetric LB error bars: $4680\pm{660}$ 
photons cm$^{-2}$ s$^{-1}$ sr$^{-1}$,
can be attributed to the halo.   
Given the extinction along our line of sight due to an
$N_{\rm{HI}}$ of 0.5 to $2.0 \times 10^{20}$~cm$^{-2}$,
the halo's intrinsic intensity is
%unrounded: $5345^{+648}_{-754}$ 
$5350^{+650}_{-750}$ to 
%unrounded: $7964^{+965}_{-1124}$ 
$7960^{+970}_{-1120}$ photons s$^{-1}$ cm$^{-2}$ sr$^{-1}$.
%However, if the emitting region
%resides above the small amount of extincting material 
%along our line of %sight,
%%line of sight extincting material, 
%then the halo's intrinsic intensity can be as large as
%$8068^{+978}_{-1138}$ photons cm$^{-2}$ s$^{-1}$ sr$^{-1}$.

We estimated the halo's \oxysix\ column density from 
absorption lines seen in
the UV spectra of extragalactic objects.  We averaged values toward the 4 
nearest sight lines
and then subtracted an estimate for the expected contribution
from the Local Bubble. 
When we used it and our intrinsic \oxysix\ intensity range and
treated 
%these values as if they derived from isothermal plasma, 
the \oxysix-bearing gas as if it were isothermal,
we found that
the electron density and thermal pressure in the halo's \oxysix-rich
gas are 
0.01 to 0.02~cm$^{-3}$ and 7000 to 10,000~K~cm$^{-3}$, respectively.
%0.01~cm$^{-3}$ and 6000~K cm$^{-3}$, respectively, if the
%\oxysix\ resides below the small amount of extincting material 
%along our line of sight and 
%0.02~cm$^{-3}$ and $10^4$~K cm$^{-3}$, respectively, if the
%\oxysix\ resides above the extincting material.

By performing a similar shadowing analysis with the \rosat\
1/4~keV data, we determined the 1/4 keV count rate attributable
to the Galactic halo along
$\ell = 278.7^{\rm{o}}, b = -47.1^{\rm{o}}$.   It
is 732 $\pm$ 142 $\times 10^{-6}$ counts s$^{-1}$ arcmin$^{-2}$, 
before accounting for line of sight extinction.
After accounting for extinction, the intrinsic R12 countrate is
$779 \pm 151$ to 
%unrounded:  $4771 \pm 927$ 
$4770 \pm 930$ counts s$^{-1}$ arcmin$^{-2}$.
%if the emission originates in front of the extincting material and
%4911 $\pm$ 953 $\times 10^{-6}$ counts s$^{-1}$ arcmin$^{-2}$ if the
%emission originates behind the small amount of extincting material
%along our line of sight.
The \oxysix\ vs 1/4~keV intensity ratio is 
%4.4 to 1.1, 
4.7 to 1.1, depending
on whether the emitting gas is beyond 
$N_{\rm{HI}} = 0.5$ or $2.0 \times 10^{20}$~cm$^{-2}$, respectively.
%extincting material is behind or in front of the
%emitting gas, respectively.   
Thus, more energy leaves the system through
the \oxysix\ resonance lines than through the \rosat\ 1/4~keV bandpass.
In contrast, the opposite is true of the local region (Local Bubble
$+$ heliospheric), where
roughly twice as much energy (at least) is radiated by the
1/4~keV emission lines than by the \oxysix\ resonance line doublet.

In order to estimate the maturity of the emitting plasma, we compared
the \oxysix\ to soft X-ray ratio with predictions for a simulated
supernova remnant at various times in its life.
Around the time that the thermal temperature throughout most of the
remnant approached the
plasma's collisional ionizational equilibrium temperature, 
the spectrum produced a 4.7 to 1 ratio, coincident with the observational
ratio assuming 
minimal extinction.   Earlier in its life, the simulated remnant 
had produced a 1.1 to 1 ratio, coincident with the observational
ratio assuming mild extinction. 
Specifically, these ratios were achieved when the SNR was
70,000 and 40,000 years old, respectively. 
We suggest that other possible hot gas structures 
would be similarly mature when they, too, produce a similar
\oxysix\ to soft X-ray intensity ratio.

Using the \oxysix\ to soft X-ray ratio, we were also able to
parameterize the hot halo plasma with a power law
temperature distribution ($dl = BT^\beta 
d\ln T~{\rm for}~10^5\,{\rm K}<T<T_{\rm 
cut}$).
Given the nominal estimates of foreground absorption
($N_{\rm HI} = 1.0 \times 10^{20}$ cm$^{-2}$) and
\oxysix\ column density 
($N_{\rm OVI} = 2.09 \times 10^{14}$ cm$^{-2}$) along our line of sight,
we found $\beta = 1.48 \pm 0.18$, $T_{\rm cut} = 10^{6.6}$ K,
and $B = 10^{-6.16 \pm 0.37}$ K$^{-\beta}$ pc.
Hot gas with a temperature between $10^5$ K and $T_{\rm cut}$ occupies
$\int dl = 10^{3.44\pm0.37}$ pc along our line of sight, but the
\oxysix-rich gas occupies a small fraction of this length.
Assuming that our line of sight is typical of high latitude sight lines,
we found the cooling rate for the halo (both sides of the
plane beyond the Local Bubble) per unit cross sectional area to be 
$dU/dt = 10^{38.73\pm0.06}$ erg s$^{-1}$ kpc$^{-2}$.
At the Sun's galactocentric radius,
the hot halo's radiative cooling accounts for $\sim70\%$ of
the energy injected into the ISM from SNe and pre-SN winds
in the galactic disk and halo.
%including SNe and pre-SN winds in the galactic disk. 
The remaining $\sim30\%$ of the injected energy must be
split between all other energy loss processes.

\acknowledgements
We appreciate K.D. Kuntz's assistance with the DIRBE corrected IRAS
data, D. Henley's comments on X-ray observations of
the Local Bubble and Galactic halo, and S. Snowden's provisions
of the digitized versions of the 
\rosat\ response curves shown in Fig. 7 of Snowden et al. (1997).
This work was funded through NASA grants NNG04GD77G 
and NNG04GD78G to the University of 
Georgia and grant NAG5-12519 to Princeton University.
This paper utilized observations obtained by the NASA-CNES-CSA 
{\it Far Ultraviolet Spectroscopic Explorer (FUSE)} 
mission operated by Johns
Hopkins University, supported by NASA contract NAS5-32985. \\

\clearpage

%The new method #2 numbers are from Shauna's Aug. 11, 2005 e-mail.
%The commented out fit method $\#2$ entries are Shauna's results for
%fitting the 1032~line independently of the 1038~line.   
%The method $\#1$ measurements were from $\sim$ July 29, 2005.  
\begin{deluxetable}{l|cc|cc}
\tablewidth{0pt}
\tablecaption{Observed Intensities and 1 $\sigma$ Statistical 
Uncertainties}
\tablehead{
\colhead{  }     
& \colhead{Night Only }
& \colhead{Night Only }
& \colhead{Day+Night  }
& \colhead{Day+Night  } \\
\colhead{  }     
& \colhead{Method $\#$1}
& \colhead{Method $\#$2}
& \colhead{Method $\#$1} 
& \colhead{Method $\#$2} \\
\colhead{  }     
& \colhead{(ph s$^{-1}$ cm$^{-2}$ sr$^{-1}$)}
& \colhead{(ph s$^{-1}$ cm$^{-2}$ sr$^{-1}$)}
& \colhead{(ph s$^{-1}$ cm$^{-2}$ sr$^{-1}$)}
& \colhead{(ph s$^{-1}$ cm$^{-2}$ sr$^{-1}$)}
} 
\startdata
% unrounded:
%\oxysix\ 1032 \AA & 3266 $\pm$ 455 & 3151 $\pm$ 535 & 3151  $\pm$ 345  
%& 3681 
%$\pm$ 420 \\
%\oxysix\ 1038 \AA & 1439 $\pm$ 378 & 2602 $\pm$ 630 & 1494 $\pm$ 282  
%& 1615 
%$\pm$ 380 \\ 
%\cartwostar\ 1037 \AA& 1547 $\pm$ 332& 1550 $\pm$ 400 & 1696 $\pm$ 246  
%& 1907 
%$\pm$ 300  \\
%1031 \AA\ feature & --             & -- & 1768 $\pm$ 260  
%& 1772 $\pm$ 265  \\ 
%
% rounded:
\oxysix\ 1032 \AA & 3270 $\pm$ 460 & 3150 $\pm$ 540 & 3150  $\pm$ 350  
& 3680 
$\pm$ 420 \\
\oxysix\ 1038 \AA & 1440 $\pm$ 380 & 2600 $\pm$ 630 & 1490 $\pm$ 280  
& 1620 
$\pm$ 380 \\ 
\cartwostar\ 1037 \AA& 1550 $\pm$ 330& 1550 $\pm$ 400 & 1700 $\pm$ 250  
& 1910 
$\pm$ 300  \\
1031 \AA\ feature & --             & -- & 1770 $\pm$ 260  
& 1770 $\pm$ 270  \\ 
\enddata
\label{table:intensityresults}
\end{deluxetable}

% For this table, the method $\#1$ measurements were from $\sim$ 
%July 29, 2005 and in it, the ``both'' 1038 \AA\ line is set to 
%match night range.
\begin{deluxetable}{l|cc|cc}
\tablewidth{0pt}
\tablecaption{Velocity with Respect to the Local Standard of Rest}
\tablehead{
\colhead{  }     
& \colhead{Night Only }
& \colhead{Night Only }
& \colhead{Day+Night  }
& \colhead{Day+Night  } \\
\colhead{  }     
& \colhead{Method $\#$1}
& \colhead{Method $\#$2}
& \colhead{Method $\#$1}
& \colhead{Method $\#$2} \\
\colhead{  }     
& \colhead{(km sec$^{-1}$)} 
&\colhead{(km sec$^{-1}$)} 
& \colhead{(km sec$^{-1}$)} 
& \colhead{(km sec$^{-1}$)}
}
\startdata
% unrounded:
%\oxysix\ 1032 \AA     &  28 &  25 $\pm$ 10      &  22  & 14 $\pm$ 10   \\
%\oxysix\ 1038 \AA     & -20 & -24 $\pm$ 28      &  -5  & -2 $\pm$ 13  \\
%\cartwostar\ 1037 \AA & -24 & -31 $\pm$ 9 	& -17  & -13 $\pm$ 7   \\
%
% rounded error bars
\oxysix\ 1032 \AA     &  28 &  25 $\pm$ 10      &  22  & 14 $\pm$ 10   \\
\oxysix\ 1038 \AA     & -20 & -24 $\pm$ 30      &  -5  & -2 $\pm$ 10  \\
\cartwostar\ 1037 \AA & -24 & -31 $\pm$ 10 	& -17  & -13 $\pm$ 10   \\
\enddata
\label{table:velocityresults}
\end{deluxetable}

% table formatted for astro-ph
\begin{deluxetable}{cc|cccc}
\tablewidth{0pt}
\rotate
\tablecaption{Physical Parameters of Halo \oxysix-rich Gas, 
Isothermal Case}
\tablehead{
\colhead{Assumed $N_{\rm H}$}
& \colhead{Transmission}
& \colhead{$I_{\rm OVI}$}
& \colhead{$N_{\rm OVI}$}
& \colhead{$n_e$} 
& \colhead{$p_{th}/k$} \\
\colhead{}
& \colhead{}
& \colhead{}
& \colhead{}
& \colhead{$\Delta l$}
& \colhead{$t_{cool}$} \\
\colhead{($10^{20}$~cm$^{-2}$)}
& \colhead{}
& \colhead{(ph s$^{-1}$ cm$^{-2}$ sr$^{-1}$)}
& \colhead{(ions cm$^{-2}$)}
& \colhead{(cm$^{-3}$)}
& \colhead{(K cm$^{-3}$)} \\
\colhead{}
& \colhead{}
& \colhead{}
& \colhead{}
& \colhead{(pc)} 
& \colhead{(years)} \\
} 
\startdata
$0.5$ & 88$\%$   & $5350^{+650}_{-750}$ & 
$2.09 \pm 0.84 \times 10^{14}$ & 
$0.0110^{+0.0046}_{-0.0047}$ & 
$6740^{+2820}_{-2860}$ \\
  &  &  &  & $69.8^{+56.8}_{-56.5}$ & $7.36^{+3.13}_{-3.08} \times 10^{6}$ \\
$1.0$ & $77\%$	 & $6110^{+740}_{-860}$	& '' & 
$ 0.0125^{+0.0052}_{-0.0053} $
& $7690^{+3220}_{-3270}$  \\
  &  &  &  & $61.1^{+49.7}_{-49.5}$ & $6.45^{+2.74}_{-2.70} \times 10^6$ \\
$2.0$ & 59$\%$ & $7960^{+970}_{-1120}$	& '' & 
$ 0.0163^{+0.0068}_{-0.0069}$ &
$10040^{+4200}_{-4260}$ \\
  &  &  &  & $46.8^{+38.1}_{-37.9}$ & $4.94^{+2.10}_{-2.07} \times 10^6$  \\
\enddata
\label{table:parameters}
\tablecomments{
Table is ``line-wrapped''; $\Delta l$ and $t_{cool}$ appear
below $n_e$ and $p_{th}/k$.}
\end{deluxetable}

\begin{deluxetable}{
r       % Parameter
c       % beta+-
c       % NH=0.5
c       % NH=1.0
c       % NH=2.0
c       % Eq. nr.
}
\tabletypesize{\small}
\tablecolumns{6}
\tablewidth{0pt}
\tablecaption{Parameters for the Temperature Power Law}
\label{pwr_law_parameters}
\tablehead{
\colhead{} & \colhead{} 
& \multicolumn{3}{c}{Foreground 
$N({\rm H~I})~(10^{20}{\rm cm}^{-2})$} & \colhead{}\\
\cline{3-5}\\
\colhead{Parameter\tablenotemark{a}} & \colhead{} & \colhead{0.5} & 
\colhead{1.0} & \colhead{2.0} & \colhead{Eq. nr.}\\
}
\startdata
$\beta$&&                                               
$1.15\pm 0.20$&         
$1.48\pm 0.18$&         $1.95\pm 0.17$& 
\protect\ref{ratio_IOVI_R12}\\[5pt]
$\log T_{\rm cut}~({\rm K})$&&                     6.9&                  
6.6&                    6.4&            \protect\ref{ratio_R67_R12}\\[5pt]
$\log p_{th}/k~({\rm cm}^{-3}\,$K)& ($\beta +$)&    $3.86\pm 0.19$&       
$3.93\pm 0.19$&         $4.08\pm 0.19$& \protect\ref{ratio_IOVI_NOVI}\\
 &                                      ($\beta  $)&    $3.85\pm 0.19$&  
$3.92\pm 0.19$&         $4.06\pm 0.19$\\
 &                                      ($\beta -$)&    $3.84\pm 0.19$&  
$3.91\pm 0.19$&         $4.05\pm 0.19$\\[5pt]
$<v^2>~({\rm km}^2{s}^{-2})$&           ($\beta +$)&    208&      
225&                    259&            \protect\ref{vel_var}\\
 &                                      ($\beta  $)&    198&       
214&                    242\\
 &                                      ($\beta -$)&    191&     
205&                    229\\[5pt]
$\log B~({\rm K}^{-\beta}{\rm pc})$&    ($\beta +$)&    $-5.37\pm 0.37$& 
$-7.12\pm 0.37$&        $-9.95\pm 0.37$&\protect\ref{NOVI_dlnT}\\
 &                                      ($\beta  $)&    $-4.25\pm 0.37$& 
$-6.16\pm 0.37$&        $-8.92\pm 0.37$\\
 &                                      ($\beta -$)&    $-3.14\pm 0.37$&
$-5.15\pm 0.37$&        $-7.96\pm 0.37$\\[5pt]
$\log L$ (pc)&                          ($\beta +$)&    $3.81\pm 0.37$& 
$3.56\pm 0.37$&         $3.36\pm 0.37$& \protect\ref{L}\\
 &                                      ($\beta  $)&    $3.62\pm 0.37$& 
$3.44\pm 0.37$&         $3.26\pm 0.37$\\
 &                                      ($\beta -$)&    $3.43\pm 0.37$& 
$3.31\pm 0.37$&         $3.18\pm 0.37$\\[5pt]
$\log N_e~({\rm cm}^{-2})$&            ($\beta +$)&    $19.46\pm 0.19$&   
$19.46\pm 0.19$&        $19.50\pm 0.19$&\protect\ref{Ne}\\
 &                                      ($\beta  $)&    $19.34\pm 0.19$&  
$19.37\pm 0.19$&        $19.43\pm 0.19$\\
 &                                      ($\beta -$)&    $19.25\pm 0.19$&  
$19.30\pm 0.19$&        $19.36\pm 0.19$\\[5pt]
$\log [dEM(T)T^{2-\beta}/d\ln T]$ &     ($\beta +$)&    $1.78\pm 0.06$&   
$0.18\pm 0.06$&         $-2.36\pm 0.06$&\protect\ref{EM}\\
$({\rm K}^{\beta-2}{\rm cm}^{-6}{\rm pc})$
 &                                      ($\beta  $)&    $2.88\pm 0.06$&   
$1.12\pm 0.06$&         $-1.36\pm 0.06$\\
 &                                      ($\beta -$)&    $3.98\pm 0.06$&   
$2.11\pm 0.06$&         $-0.42\pm 0.06$\\[5pt]
$\log(dU/dt)~({\rm erg~s}^{-1}{\rm kpc}^{-2})$ &
                                        ($\beta +$)&    $38.69\pm 0.06$&  
$38.76\pm 0.06$&        $38.72\pm 0.06$&\protect\ref{dEdt}\\
 &                                      ($\beta  $)&    $38.73\pm 0.06$&  
$38.73\pm 0.06$&        $38.78\pm 0.06$\\
 &                                      ($\beta -$)&    $38.77\pm 0.06$&  
$38.76\pm 0.06$&        $38.79\pm 0.06$\\
 &
\enddata
\tablecomments{
%Note: 
Errors appended to the listed quantities are the formal
errors that arise from uncertainties in the measured quantities
only.  They do not include systematic errors caused by inaccuracies
in atomic physics parameters, element abundances, the assumption of
isobaric conditions, or deviations from our temperature power-law
representation.}
\tablenotetext{a}{($\beta +$) indicates value at the largest 
limit of $\beta$, 
($\beta$) indicates
the value at the preferred value of $\beta$, and ($\beta -$) 
indicates the value 
at the smaller
limit for $\beta$.}
\label{pwr_law_parameters}
\end{deluxetable}

\clearpage

\end{document}